\documentstyle[12pt,aaspp4,psfig,flushrt]{article}

\newcommand{\ctan}{$^{13}$C($\alpha$,n)$^{16}$O}
\newcommand{\ctanb}{$^{13}$C($\alpha$,n)$^{16}$O~}
\newcommand{\nean}{$^{22}$Ne($\alpha$,n)$^{25}$Mg}
\newcommand{\ndd}{$^{22}$Ne}

\newcommand{\msb}{$M_{\odot}$~}
\newcommand{\ms}{$M_{\odot}$}
\newcommand{\cd}{$^{12}$C}
\newcommand{\cdb}{$^{12}$C~}
\newcommand{\ct}{$^{13}$C}
\newcommand{\ctb}{$^{13}$C~}
\newcommand{\s}{$s$}

\begin{document}

\title{NUCLEOSYNTHESIS AND MIXING ON THE ASYMPTOTIC GIANT BRANCH.\\
 III. PREDICTED AND OBSERVED $s$-PROCESS ABUNDANCES}

\vspace{0.5cm}

\author{Maurizio Busso\altaffilmark{1},
Roberto Gallino\altaffilmark{2}, David L. Lambert\altaffilmark{3},\\
Claudia Travaglio\altaffilmark{4}, and Verne V.
Smith\altaffilmark{5}}

\vspace{8cm}

\affil{1. Osservatorio Astronomico di Torino, Torino, I-10025 Pino
Torinese, Italy, busso@to.astro.it}

\affil{2. Dipartimento di Fisica Generale, Universit\`a di
Torino, Via P. Giuria 1, I-10125 Torino, Italy,
gallino@ph.unito.it}

\affil{3. Department of Astronomy, University of Texas at Austin,
TX 78741, USA, dll@astro.as.utexas.edu}

\affil{4. Max-Planck Institut f\"ur Astronomie, K\"onigsthul 17,
D-69117 Heidelberg, Germany, claudia@mpia-hd.mpg.de}

\affil{5. Department of Astronomy, University of Texas at El Paso,
TX 79968, USA, verne@barium.physics.utep.edu}

\newpage

\begin{abstract}
We present the results of $s$-process nucleosynthesis
calculations for Asymptotic Giant Branch (AGB) stars of different
metallicities and different initial stellar masses (1.5 and 3
\ms) and comparisons of them with observational constraints from high
resolution spectroscopy of evolved stars over a wide metallicity range.
The computations were based on previously published stellar
evolutionary models that account for the third dredge up
phenomenon occurring late on the AGB. Neutron production is
driven by the $^{13}$C($\alpha$,n)$^{16}$O reaction during the
interpulse periods in a tiny layer in radiative equilibrium at
the top of the He- and C-rich shell. The neutron source \ctb is
manufactured locally by proton captures on the abundant \cd; a
few protons are assumed to penetrate from the convective envelope
into the radiative layer at any third dredge up episode, when a
chemical discontinuity is established between the convective
envelope and the He- and C-rich zone.  A weaker neutron release
is also guaranteed by the marginal activation of the reaction
$^{22}$Ne($\alpha$,n)$^{25}$Mg during the convective thermal
pulses.  Owing to the lack of a consistent model for \ctb
formation,  the abundance of \ctb burnt per cycle is allowed to
vary as a free parameter over a wide interval (a factor of 50).
The $s$-enriched material is subsequently mixed with the envelope
by the third dredge up, and the envelope composition is computed
after each thermal pulse. We follow the changes in the
photospheric abundance of the Ba-peak elements (heavy $s$, or
`hs') and that of the Zr-peak ones (light $s$, or `ls'), whose
logarithmic ratio [hs/ls] has often been adopted as an indicator
of the $s$-process efficiency (e.g. of the neutron exposure). Our
model predictions for this parameter show a complex trend versus
metallicity. Especially noteworthy is the prediction that the
flow along the $s$ path at low metallicities drains the Zr-peak
and Ba-peak and builds an excess at the doubly-magic $^{208}$Pb,
at the termination of the $s$ path. We then discuss the effects
on the models of variations in the crucial parameters of the \ctb
pocket, finding that they are not critical for interpreting the
results.

The theoretical predictions are compared with published
abundances of $s$ elements for AGB giants of classes MS, S, SC,
post-AGB supergiants, and for various classes of binary stars,
which supposedly derive their composition by mass transfer from
an AGB companion. This is  done for objects belonging both to the
Galactic disk and to the halo. The observations in general
confirm the complex dependence of neutron captures on metallicity.
They suggest that a moderate spread exists in the abundance of \ctb that
is burnt in different stars. Although additional observations are
needed, it seems that a good understanding has been achieved of
$s$-process operation in AGB stars. Finally, the detailed
abundance distribution including the light elements (CNO)
 of a few $s$-enriched stars at different
metallicity are examined and satisfactorily reproduced by model
envelope compositions.

\end{abstract}


\section{Introduction}

In this paper we present an interpretation of the observed
$s$-element abundances in evolved red giants in the light of
current models for the nucleosynthesis occurring in low mass
stars. The red giants studied here are primarily those powered by
hydrogen and helium burning in two shells located above a
degenerate C-O core and are called Asymptotic Giant Branch (AGB)
stars, see e.g. Iben \& Renzini (1983). Late on the AGB, they
undergo recurrent thermal instabilities of the He-shell (thermal
pulses, TP), where partial He burning occurs convectively over
short periods of time, sweeping the whole region that lies
between the H and the He shells (hereafter, the He intershell).
The stars undergoing such processes are semiregular or Mira
variables showing the short-lived Tc in their spectra (Merrill
1952). They belong to the spectral types MS, S, SC, or C(N) and
are called {\it intrinsic} TP-AGB stars (Lambert 1985; Smith \&
Lambert 1990, hereafter SL90; Plez, Smith, \& Lambert 1993).
Strong and variable stellar winds, with mass loss rates
 $dM/dt$ from $10^{-7}$ up to $10^{-5}$ \ms/yr, progressively
reduce their envelopes, leaving behind a degenerate C-O core,
which will eventually become a white dwarf. During the TP-AGB
phase, material from the He intershell, enriched in \cd,
$^{22}$Ne, and \s-process elements, appears at the surface,
thanks to repeated downward extensions of the envelope convection
(third dredge up, TDU). In other cases, the star that was the
site of the \s-element synthesis appears to have left the TP-AGB
phase, ejecting most of its envelope and exposing the hot layers
immediately above the H shell. Such a star appears as a post-AGB
carbon-rich and \s-element-rich supergiant of classes A, F, or G
(Van Winckel \& Reyniers 2000).

Further constraints on neutron-capture nucleosynthesis in the
Galaxy come from binary systems, in which the more massive
component, while on the TP-AGB phase, has transferred part of its
envelope onto a lower mass companion, now seen to be enriched in
$s$-elements. It is  therefore this companion ({\it extrinsic}
AGB) that becomes the object of our study (see Jorissen \& Mayor
1988; Han et al. 1995; Jorissen et al. 1998 for a treatment of
mass transfer and Ba-star masses, and Smith 1984, Luck \& Bond
1991a,b, Luck, Bond, \& Lambert 1990 for details on the resulting
abundances). In this group of stars we find the classical Ba II
giants, and other sources in which the secondary component has in
its turn evolved to the early AGB phases, so that it can be
distinguished from `normal' MS, S, SC and C(N) stars only by the
absence of Tc. The group of extrinsic AGBs also contains several
classes of CH-strong dwarfs, subgiants and giants at various
metallicities. These are low luminosity stars with $s$-process
enrichments that are most probably not spectroscopic binaries,
e.g., the N-rich subdwarf HD~25329 (Beveridge \& Sneden 1994), and
HD~23439A and B (Tomkin \& Lambert 1999). In these cases, it is
presumed that their natal clouds were contaminated by ejecta from
AGB stars. This is e.g. the case of the $s$ enrichment in red
giants belonging to the globular cluster $\omega$ Cen (Smith et
al. 2000).

As far as nucleosynthesis is concerned, current studies ascribe
neutron captures to the activation of the reactions
$^{13}$C($\alpha$,n)$^{16}$O and $^{22}$Ne($\alpha$,n)$^{25}$Mg
(Iben \& Renzini 1982; Hollowell \& Iben 1988; Gallino et al.
1988; K\"appeler et al. 1990; Busso, Gallino, \& Wasserburg 1999,
hereafter BGW99). The same reactions may also affect the
nucleosynthesis of lighter nuclei, like fluorine, sodium, and the
unstable $^{26}$Al (Mowlavi \& Meynet, 2000; Goriely \& Mowlavi
2000; Mowlavi 1999).

The principal goal of this paper is to verify the ability of
current forms of AGB models to account for the observed abundance
distributions in evloved stars. In \S~2 we discuss the general
features of our models and the choices made for the
parameterization of the neutron source \ct. We also comment on
the major effects on \s-process nucleosynthesis introduced by
varying the initial metallicity. A comparison of observed heavy
element abundances of intrinsic and extrinsic $s$-enriched stars
with model predictions follows in \S~3. Detailed fits to the
observed compositions including the light elements of three
selected stars are sketched in \S~4. Finally, in \S~5 we
summarize the main conclusions that can be drawn from such an
analysis.

\section{The Adopted Models for Stellar Evolution and Nucleosynthesis}

\subsection{Indices of $s$-processing}

Abundance analyses of stars enriched by $s$-processing may often
provide quantitative information for only a few neutron rich
elements. This has led to simple characterizations of the
overabundances. Following Luck \& Bond (1991a,b), it is common to
monitor the \s-process efficiency through the relative abundances
of the $s$ elements at the Ba-peak (collectively indicated as
`heavy $s$', or `hs') with respect to those at the Zr-peak
(indicated as `light $s$' or `ls'). Indeed, those nuclei are
placed at neutron magic numbers N $=$ 82 and 50, are mainly of
\s-process origin and act as bottlenecks for the \s-process path
because of their low neutron capture cross  sections.
Consequently, for relatively low $s$-process efficiencies, the
neutron flux mainly feeds the nuclei at the Zr-peak, while for
higher exposures the Ba-peak species are favored (see \S~ 3.1 for
a deeper analysis). The average logarithmic ratio normalized to
solar, [hs/ls], defined as [hs/ls] = $\log~$(hs/ls)$_*
-\log~$(hs/ls)$_\odot$,  has been extensively  used as a measure
of the  neutron capture efficiency in building up the heavy
\s~elements and proved useful for the interpretation of the
Galactic disk stars (Busso et al. 1992, 1995, hereafter Paper I
and Paper II, respectively). The choice of the specific elements
to consider in the average `hs' and `ls' abundances varies from
author to author and depends on the quality of the spectra
available. The definition adopted by Luck \& Bond (1991a,b)
included Sr, Y, Zr in the `ls' group and Ba, La, Nd, Sm in the
`hs' one. In presenting the model results we instead consider Y
and Zr as `ls'. We generally exclude Sr, despite the fact that it
is mainly of $s$-process origin, because it has been measured
precisely only in a minority of the sample stars. For the `hs'
group, we include La and Nd, which are present with several lines
in most sample stars. The results would not be significantly
changed if we adopted the same choice as Luck \& Bond (1991a,b).
Whenever possible, we follow the same rules in selecting the
observations. There are however cases in which this is not
possible. One is represented by the star CS 22898-027 (McWilliam
et al. 1995) where `ls' can be defined only through Sr (Y and Zr
were not measured). For `hs' the situation is more complex. There
are cases in which only one element of the couple La, Nd was
measured, in other cases only estimates of the Ba abundance
exist, etc. The stars affected by such problems are commented
explicitly later, in order to make clear which choices were made
(see discussion of Table 2). These limitations increase slightly
the uncertainties in the constraints.

In addition, the degree of $s$-process enrichment is of interest.
We consider the logarithmic ratios, normalized to solar, [ls/Fe]
and [hs/Fe] as measures of the enrichment. For the intrinsic
stars, both indices increase with the number of TDUs but the
ratio [hs/ls] after a relatively few TDUs approaches an
asymptotic value. In the case of the extrinsic stars, the [ls/Fe]
and [hs/Fe] are dependent on the values of these indices in the
AGB envelope during mass transfer and on the degree of dilution
occurring in the extrinsic star on receipt of the mass or
subsequently.

For a few stars, a quite extensive array of heavy elements have
been measured, and, in these cases, theoretical models may be
tested for their ability to reproduce the suite of elemental
abundances, a finer test than that possible using indices `ls'
and `hs'.

\subsection{General Characteristics of $s$ Processing in AGB Stars}

As mentioned, the main neutron source allowing $s$-processing to
take place in AGB stars is the reaction \ctan. Its activation
requires the penetration of a small amount of protons from the
envelope into the He intershell (i.e. the region between the H
and He shells), most likely occurring during the post-flash
luminosity dip, when the H shell is extinguished and TDU can
occur (Iben \& Renzini 1983). Studies on the required mixing
mechanisms have made use of diffusive or hydrodynamical
simulations (Iben \& Renzini 1982; Hollowell \& Iben 1988; Herwig
et al. 1997; Herwig 2000; Cristallo et al. 2001). They are still
a matter of debate, though a consensus is emerging that partial
mixing must indeed occur. At least in some cases it may also be
affected by rotational shears (Langer et al. 1999), but a
self-consistent model is still not possible in stellar codes. Due
to these uncertainties, the amount of $^{13}$C that forms in a
{\it pocket} at the top of the He intershell and its profile in
mass must be represented by free parameters. Introducing a range
of \ctb abundances at a given metallicity proved to be effective
for several purposes (see, e.g., Gallino et al. 1998, hereafter
G98; Travaglio et al. 1999, hereafter T99, Goriely \& Mowlavi,
2000).

The alternative to the \ct~neutron source  is the \nean~
reaction, which is triggered
 when the temperature $T$ exceeds 3$\times$10$^8$ K, but this reaction
does not play a dominant role in AGB stars of $M$ $\le$ 3 $-$ 4
\msb (hereafter Low Mass Stars, or LMS). In fact in LMS the
maximum temperature at the bottom of TPs, though gradually
increasing with the pulse number, barely reaches the above
mentioned value. Despite the very low neutron exposure associated
with its activation in LMS, the peak neutron density generated by
the \ndd~ source is rather high (up to $n_n = 10^{10}$ cm$^{-3}$)
and affects many branchings along the \s~path that are sensitive
to the neutron density and/or to temperature (K\"appeler et al.
1990; Arlandini et al. 1999). Notice that the final modifications
on branching-dependent isotopes are important, since the
\s-processed and partially diluted matter is irradiated
repeatedly and therefore keeps a partial memory of all the
previous high-temperature phases.

In more massive AGB stars (5 $<$ $M$/\msb $<$ 8, hereafter
Intermediate Mass Stars, or IMS), \ndd~ burns efficiently through
$^{22}$Ne($\alpha$,n)$^{25}$Mg and
$^{22}$Ne($\alpha$,$\gamma$)$^{26}$Mg reactions, in roughly equal
proportion, since the maximum bottom temperature in TPs is $T \ge$
3.5 $\times$ 10$^8$ K (Iben 1975; Truran \& Iben 1977). As a
consequence, a significant neutron exposure is made available.
Simultaneous production of $^{25}$Mg and $^{26}$Mg occurs such
that these isotopes are expected to be enhanced in the
photospheres of AGB stars of intermediate mass. Moreover, the
high peak neutron density ($n_n$ $>$ 10$^{11}$ cm$^{-3}$) induces
a considerable production of the few neutron-rich nuclei involved
in those $s$-process branchings sensitive to the neutron density,
such as $^{86}$Kr, $^{87}$Rb, and $^{96}$Zr. Their abundance can
be enhanced even more than for the $s$-only species. Moreover,
IMS may suffer from the so-called hot bottom burning process
(HBB) in the deep convective envelope, which would decrease the
photospheric \cd/\ctb ratio and enhance the abundance of Li. It
would also reduce the C/O ratio, thus preventing the star from
becoming C-rich for most of its AGB phase (e.g., Lattanzio \&
Forestini 1999).

The observational evidence so far available points toward a relatively low
peak neutron density in AGB stars ($n_n$ $<$ 10$^{8}$ cm$^{-3}$), as
derived from the Sr/Rb abundance ratio and in
a few cases from the Zr isotopic mix (Lambert et al. 1995). This can be
interpreted as an indication that most intrinsic and extrinsic AGB stars are
of rather low  mass (Busso et al. 1988).
The fact that AGB  stars of intermediate mass are
rather rare is then confirmed by the roughly solar isotopic ratio measured
for Mg in MS and S giants (see e.g. Clegg, Lambert, \& Bell 1979; Blanco,
McCarthy, \& Blanco 1980; McWilliam \& Lambert 1988).

\subsection{The Updated reference Models}

Stellar models  were computed using the FRANEC code (Straniero et
al. 1997; G98) and spanned the metallicity range from solar down
to 1/20 solar (1.5 \ms) and from solar down to 1/3 solar (3 \ms).

Our nucleosynthesis predictions, covering the metallicity
interval from [Fe/H] = +0.3 down to [Fe/H] = $-$3.6, were
calculated as post-process runs, extrapolating the stellar
parameters from the mentioned coarser grid of full evolutionary
models. The differences in the physical structure of the He
intershell are found to be small, even for models with large
differences in the initial metallicities (see also Boothroyd \&
Sackmann 1988, 1999). The parameter most heavily affected by the
metal content is the TDU, which is found to increase in
efficiency for decreasing [Fe/H] values. For these reasons, the
amount of dredge up at very low metallicities may be somewhat
underestimated in our extrapolations, and TDU can occur also for
masses lower than actually found. Mass loss was considered
through the Reimers (1975) formula, varying its free parameter to
obtain a wide range of parameterizations. The values considered
in this paper are $\eta$ = 0.3 (1.5 \ms), $\eta$ = 1.5 (3 \ms).
With the FRANEC code and its assumptions for convective mixing,
TDU is found to cease when the residual envelope mass becomes
smaller than about 0.5 \ms, while thermal pulses are still going
on. Given the uncertainties of the mixing procedure in stellar
models, and of the mass loss rates, we cannot claim that this
last finding is a common property of real AGB stars. However, on
this point the different stellar codes tend to agree (see also
Lattanzio \& Karakas 2001). Should this correspond to a real
physical property, then it would also imply that stars with too
small envelope masses (hence also with too small {\it initial}
masses) cannot undergo TDU, and cannot contribute to the chemical
enrichment of the Galaxy (T99). The minimum initial mass at which
TDU is found with the FRANEC code is 1.5 \msb for a solar
composition, and decreases slightly for decreasing metallicity:
however the uncertainty is high and the actual values should be
tuned through comparisons with observations (see also \S~3.2 and
the discussion of Figure 7). Exceptional conditions have probably
to be invoked to explain objects with unusual $s$ enrichments, up
to 1 order of magnitude higher than the maximum defined by common
AGB stars: see e.g. the cases of F Sge (Gonzalez et al. 1997),
the Sakurai's object (Asplund et al. 1999) and in general the
class of R CrB stars (Asplund et al. 2000). We also recall that
in Paper I we analysed which would be the consequences of a
prolonged interruption in the TDU process. We showed that, in
this case, some MS or S stars without Tc might actually be
intrinsic, Tc having decayed due to lack of refurbishing from the
He intershell.

The first published models for \ctb burning in AGB stars assumed
that the neutrons were released from the reaction
$^{13}$C($\alpha$,n)$^{16}$O in convective conditions, after the
\ctb pocket was ingested by the next TP (Hollowell \& Iben 1998;
K\"appeler et al. 1990). The resulting \s-process nucleosynthesis
was analysed  to understand abundances in AGB stars of the
Galactic disk (Paper I and Paper II). In such calculations, a
repeated neutron exposure was achieved thanks to partial
overlapping of material cycled through several TPs. Though more
complex  than in the simple original sketch by Ulrich (1973), the
\s-process mechanism could still be approximated by an
exponential distribution of neutron exposures. We recall that the
classical analysis of the {\it main} component  represented by
the solar-system \s-process abundances in the atomic mass range
$A$ $=$ 85 $-$ 208 (e.g., K\"appeler, Beer \& Wisshak 1989),
assumed a distribution function of neutron exposures
$\rho$($\tau$) $\sim$ exp($-$$\tau/\tau_0$), where $\tau$ is the
time integrated neutron flux, $\tau = \int n_n v_{th}dt$,
$v_{th}$ being the thermal velocity and $n_n$ the neutron density
(Seeger, Fowler, \& Clayton 1965). The parameter $\tau_0$ is
called {\it mean} neutron exposure. According to Arlandini et al.
(1999), the main component is best reproduced with $\tau_0$ $=$
(0.296~$\pm$~ 0.003) ($kT$/30 keV)$^{1/2}$. As a matter of fact,
previous AGB stellar models already showed that different
\s-element distributions were obtained by varying the amount of
\ctb in the pocket and/or the initial metallicity.
 In this respect, the  observed spread of [hs/ls]
values in Galactic disk  MS and S stars was explained
quantitatively by saying that they imply values of the {\it mean}
neutron exposure $\tau_0$ (expressed at 30 keV) in the range 0.2
-- 0.4 mbarn$^{-1}$.  CH and Ba stars
were found to extend the range of mean neutron exposures further,
up to $\tau_0$ (30 keV) $\sim$ 0.8 $-$ 1.0 mbarn$^{-1}$  (Paper II).

A reanalysis of AGB evolution was recently pursued by various
authors (see e.g. Herwig et al. 1997; Herwig, Sch\"onberner, \&
Bl\"ocker 1998; Frost et al. 1998; Frost, Lattanzio \& Wood
1998). We shall follow here the work by Straniero et al. (1997).
They showed that all $^{13}$C nuclei present in the pocket are
consumed locally in radiative conditions (over a time interval of
a few 10$^4$ yr) when these layers are heated up to 10$^8$ K
before the next convective instability sets in. The rate of the
\ctanb reaction (Denker et al. 1995) is rather uncertain at the
relevant temperatures ($T$ $\sim$ 0.9 $\times$ 10$^8$ K), but
 we expect
complete exhaustion of \ctb in the radiative interpulse phase for
all reasonable values of the rate (Arlandini et al. 1999). In
such conditions, the neutron fluxes are characterized by a rather
low neutron density ($n_n$ $\le$ 1 $\times$ 10$^7$ n/cm$^{3}$),
as a consequence of the relatively low temperature at which \ctb
burns. The effect of the recurrent convective instabilities is
nevertheless of the highest importance, since they dilute the
\s-enriched material of the \ctb pocket over the whole He
intershell, allowing subsequent neutron-capture episodes to occur
over a mixture containing fresh iron-seed nuclei and material
already $s$-processed in the previous cycles. The ensuing chemical
homogeneization of the He intershell also allows the neutron-rich
nuclei to be mixed with the envelope by TDU.

The fact that \ctb burns radiatively in the interpulse period
makes the \s~process more complex than previously assumed.
Further complications are due to the fact that the mass involved
in each thermal pulse decreases with time, as the core mass
increases, while the  overlapping factor $r$ between adjacent
pulses decreases with pulse number. The ensuing neutron exposure
is much closer to a superposition of few single irradiations than
to an exponential distribution of neutron exposures (see G98 for
further details).

For the sake of our comparisons with observed stars,
 we  adopt a wide range of \ctb abundances in
the pocket. We include the case indicated as standard (ST) by G98,
corresponding to 4$\times$10$^{-6}$ \msb of \ct, and values
scaled upward (by a factor 2) and downward (by factors 1.5, 3, 6,
12 and 24) with respect to that choice. The ST case was so named
because this amount of \ct~ for AGB stars in the mass range 1.5
$-$ 3 \msb at [Fe/H] $=$ $-$0.3  appears  to explain the main
(solar) component of the $s$ process (Arlandini et al. 1999).
Notice that the same result can be obtained by varying
simultaneously the metallicity and the amount of \ctb in the
pocket provided the ratio \ct/Fe remains constant;  this ratio
essentially fixes the number of neutrons available per heavy
seed. It is, however, clear that both the particular $s$-element
distribution achieved in the Sun and the spread of $s$-process
abundances observed in chemically unevolved stars  at each
metallicity (i.e. stars that have maintained their initial
composition) should be considered as the result of a complex
chemical evolution of the Galaxy, through the astration mechanism
over various generations of AGB stars of different masses,
metallicities, and \s-process efficiencies.

\subsection{Metallicity effects in the models}

For each choice of the \ctb pocket and  stellar mass, the
$s$-process efficiency in the He intershell varies according to
the initial metallicity. Figures 1 and 2 show the overproduction
factors of stable elements from Cu to Bi in the material
cumulatively mixed from the He intershell to the envelope. The
results refer to the ST choice for the \ctb pocket for AGB models
of 1.5 \ms, metallicity  varying from [Fe/H] = +0.3 down to
[Fe/H] = $-$0.6 (Figure 1) and from [Fe/H] = $-$1 down to [Fe/H] =
$-$3 (Figure 2). In these figures, the elements represented by
bold symbols owe more than 50\% of their solar abundance to the
$s$ process, according to the analysis by Arlandini et al.
(1999). The remaining elements are indicated as crosses. The
elements used here to build the `hs' and `ls' integrated
abundances are indicated explicitly in the third panel of Figure
1. In the $s$-processed material mixed with the envelope, the
unstable nuclei with  half-life longer than an interpulse period
(typically between 10$^4$ and 10$^5$ yr in LMS) but shorter than
10$^9$ yr, have been allowed to decay, as expected for extrinsic
AGB stars. For comparison, the abundances before such decays are
shown as open symbols (squares for the mainly-$s$ elements). The
isotopes involved  are the pairs $^{93}$Zr $-$ $^{93}$Nb,
$^{99}$Tc $-$  $^{99}$Ru, $^{107}$Pd $-$ $^{107}$Ag, $^{135}$Cs
$-$ $^{135}$Ba, $^{205}$Pb $-$ $^{205}$Tl. It is clear from the
figures that the element distribution is strongly dependent on
the initial composition. They show how the light $s$ elements
belonging to the Zr-peak are preferentially produced at nearly
solar metallicities.  At lower metallicities, like those  of the
old Galactic disk (e.g. $-$0.8 $\le$ [Fe/H] $\le$ $-$0.6), the
Ba-peak elements become dominant. This is so because the neutron
exposure increases with decreasing iron seed abundances. For the
most metal-poor stars (Figure 2) the flow along the $s$-process
path accumulates on $^{208}$Pb, the doubly-magic nucleus at the
termination point of the \s-path, and partly on $^{209}$Bi.
Similar trends are observed for the other choices of \ctb in the
pocket.

It is from the averaging of very different contributions, like
those illustrated in Figures 1 and 2, that Galactic chemical
evolution generates the heavy $s$-elements in the solar system
(T99). The trends of these figures show that a progressively
increased production of Pb is to be expected at the surface of
AGB stars with decreasing [Fe/H], at the expense of both the
Zr-peak (ls) and the Ba-peak (hs) elements. Therefore, for halo
stars, all criteria based solely on these last two parameters
lose their meaning. AGB stars of low metallicity are the major
producers of Galactic Pb (G98; Travaglio et al. 2001), and
represent the astrophysical site of the so-called {\it strong}
component advanced by the classical analysis (Clayton \& Rassbach
1967).

\subsection{Surface abundances and their dependence on metallicity}

Using the AGB models with mass loss prescriptions described
above, we computed the envelope compositions for each stellar
mass and metallicity after each TDU episode, from the first
occurrence of mixing with the envelope (after a limited number of
TPs, see Straniero et al. 1997)  to the end of the AGB phase.
This allows us to build a large data bank of model abundance
distributions, evolving pulse after pulse, characterized by three
different parameters (initial mass of the parent star, abundance
of \ctb in the pocket, metallicity).

In Figures 3 (a, b, c) and 4 (a, b, c), we show predicted values
of [hs/ls], [ls/Fe], and [hs/Fe] in the photosphere at the last
TDU episode for the 1.5 and 3 \msb stellar models. The non-linear
trends displayed by the plots reveal the complex dependence on
metallicity of the integrated abundances `ls' and `hs'. This is
due to the neutron exposure, which increases with decreasing
abundance of the iron seed nuclei. To illustrate this, let us
follow the ST case. Starting at metallicities higher than solar
and moving toward lower [Fe/H] values, a minimum with [hs/ls] =
$-$0.54 is first reached at [Fe/H] = 0.2 (the Zr-peak elements
are more abundantly produced for moderate exposures, hence for
relatively high metallicities). Then a maximum with [hs/ls] = 1.2
follows at [Fe/H] = $-$1. Finally, for even lower metallicities,
the $s$-process flow extends beyond the   Zr-peak and Ba-peak
nuclei to cause an accumulation at $^{208}$Pb, and [hs/ls]
declines again. Reducing the amount of \ctb burnt has the effect
of displacing the maximum and the minimum of the curve toward
lower metallicities. Note that the maxima reach values in [hs/ls]
up to 1.2, consistently higher than in all previous expectations.
The minima reach down to $-$0.7, at a [Fe/H] value depending on
the \ctb choice.

At extremely low [Fe/H] values, the scarcity of Fe nuclei becomes
important. Here their role as  seeds for the $s$-process
 is replaced  partly by lighter
(intermediate atomic mass) nuclei, whose  abundances remain high
due both to the known enhancement of $\alpha$-rich elements  in
halo stars and to the fact that they are syntesised from primary
\cdb generated in the He intershell and mixed with the envelope by
TDU. When this \cdb is transformed into $^{14}$N by the H-burning
shell, it provides more fuel for the chain
$^{14}$N($\alpha$,$\gamma$)$^{18}$F($\beta^+\nu$)$^{18}$O($\alpha$,$\gamma$)$^{22}$Ne
early in the TP. The nuclei that are progeny of \ndd, usually
considered merely as neutron filters (or `poisons') at higher
metallicities, now assume a more complex role. Neutron captures
on them generate a small leakage that crosses the iron peak, thus
allowing the $s$ processing on heavy isotopes to continue.

Due to the above phenomena, the  $s$ nuclei cannot be considered
as being simply of `secondary' nature. It is true that they
require previously built seeds to be formed, but  at
moderately low metallicities their secondary behavior is masked
by the increase in the neutron exposure, on which the resulting
$s$-process abundances strongly depend (Clayton 1988). At very
low [Fe/H] values part of the seeds feeding them are primary, so
that they achieve a trend that only slightly decreases for
decreasing [Fe/H].

\subsection{Sensitivity of the results to model parameters of \ctb burning}

Neutrons released by the \ctb neutron source are captured locally
in the  \ctb pocket. Only after their
ingestion in the next TP are the $s$ enriched layers of the \ctb pocket
 mixed together and spread over the (now
convective) He intershell. To better understand how this complex
mechanism occurs, and to study its sensitivity to input
parameters, we consider in some detail the behavior of
\s-processing in the various layers of the \ctb pocket together
with the effects of mixing and dilution operated by TPs. For any
\ctb neutron burst, the starting composition of Fe seeds and
heavy elements in the pocket is not the initial one, but rather
the composition left behind by the operation of previous TPs (see
Figure 5), already highly enriched in heavy elements.

This is illustrated schematically in Table 1, where the evolution
of three representative isotopes of the  Zr, Ba, and Pb peaks are
followed for the model of an AGB star of 1.5 \ms~ and solar
metallicity, adopting the ST choice for the \ctb pocket. The
relevant zones of the stellar structure are shown in Figure 5. For
illustration  let us consider the layers in the pocket of initial
abundance $X$(\ct) = 0.003 and 0.004. Column 1 gives the TP
number (with TDU) that precedes the corresponding \ct-formation
episode, column (2) indicates the zone to which the abundances
refer: layer 1 and layer 2, as indicated in the figure, then the
average in the \ctb pocket at the end of the interpulse phase
(zone 3 in the figure), and finally the average at the end of the
next TP (labeled as `4' in the figure). The composition of this
last region includes the effects of dilution over the extension
of the convective He intershell, with material from upper layers
containing ashes of H-shell burning, and material from lower
layers already $s$-processed in the previous TP cycles. During
each TP one has to take into account also the small neutron burst
by the $^{22}$Ne source,  but  this affects only marginally the
production of~$^{94}$Zr, $^{138}$Ba, and $^{208}$Pb. Column 3
gives the mass (in solar units) affected by the next convective
TP, column (4) gives the neutron exposure $\delta$($\tau$) (in
mbarn$^{-1}$), where $\delta$($\tau) = \int n_n v_{th}dt$ is
integrated over the interpulse period. Typical thermal conditions
during \ctb consumption correspond to $kT$ $=$ 8 keV. Column (5),
(6), and (7) give the production factors of $^{94}$Zr,
$^{138}$Ba, $^{208}$Pb relative to solar (Anders \& Grevesse
1989). Although the three selected isotopes are not of purely-$s$
origin, they owe most of their production to the $s$-process.
Following their synthesis as a function of pulse number in the two
selected zones, we note how in the first cycle a small difference
in $\delta$($\tau$) in the two regions gives rise to a huge
difference in their production factors. This lets the highly
non-linear trend of the $s$-process efficiency to show up.
However, the ratio of the production factors between the two
zones is progressively smoothed in the subsequent TPs. For
$^{94}$Zr/$^{94}$Zr$_{\odot}$ it goes from 0.21 in the first
\ct-burning episode to 0.66 in the 5$^{\rm th}$ and to 0.73 in
the 20$^{\rm th}$. For $^{138}$Ba/$^{138}$Ba$_{\odot}$ the range
is from 0.47 to 0.55 and to 0.64, respectively. (See G98 for a
discussion of asymptotic conditions reached in advanced TPs).
Concerning $^{208}$Pb, the production of this isotope at the
termination of the $s$-process path grows with pulse number more
slowly than the nuclei at the other two $s$-process peaks.

The neutron exposures listed in column (4) show a slight decline
with increasing pulse number in both selected layers. This
smoothing effect is due to the repeated averaging of the
$s$-process products operated by the TPs, which dilute the
$s$-enriched layers in the way described above, over a stellar
zone that contains about twenty times more mass than the one
where neutrons are produced. Due to these extreme and repeated
dilutions, even large variations in local details of the pocket
have only small effects on the final distribution of the $s$
elements, as our calculations demonstrate. In fact, we repeated
the whole series of nucleosynthesis computations for different
metallicities just assuming, for each choice of the total \ctb
concentration, the simplest possible profile, i.e. a constant
abundance throughout the pocket. In this exercise a mass fraction
$X_{13}$ = 0.004 represents the "smoothed" ST case. The resulting
[hs/ls] trend versus metallicity is shown in Figure 6. When
comparing it with the previous behavior (Figure 3a), it
immediately appears that the two sets of results are almost
indistinguishable. This gives us confidence in the results, even
in the absence of a self-consistent model for the formation of
the \ctb pocket.

\section{Observations of $s$-process Enrichment in Disk and Halo Stars}

\subsection{Unevolved Stars}

In our comparisons with compositions of $s$ enriched stars, we
involve both intrinsic and extrinsic AGBs. Use of indices such as
[ls/Fe], [hs/Fe], and [hs/ls] calls for an assumption about the
initial values of them. It is common to assume that the
star, while on the main sequence, had [ls/Fe] = [hs/Fe] = [hs/ls]
=0. These initial conditions are also adopted in the theoretical
calculations. Given the increasing quality of the observational
data and the sophistication of the models it is worthwhile to
review the evidence on heavy element abundances in disk and halo
stars.

 Intrinsic, that is single AGB and post-AGB
stars, build up their $s$ enrichments as they experience
successive TDUs. One predicts the enrichment and [hs/ls] to evolve as the
AGB evolves. This evolution  should be observable through abundance
analyses of a large sample of intrinsic stars.
 Extrinsic stars gain their $s$ enrichments by
 mass transfer from an AGB companion. Then, the measured [$s$/Fe]
refers to the average $s$-processed material of the AGB star after
dilution by mixing with the envelope of the presumed unevolved
companion that is now the extrinsic, $s$-enriched star. If, as is
often the case, considerable $s$-processed material is transferred, the initial
composition of the star is effectively irrelevant and the
observed [hs/ls] is the one of the material transferred to the
secondary component. Again, predictions for [$s$/Fe] and [hs/ls]
are testable using compositions of extrinsic stars.

Enrichments are quoted relative to the solar composition scaled
to the metallicity of the star. Hence [ls/Fe] denotes the
enrichment provided that [ls/Fe] = 0 for an unevolved star of
that metallicity [Fe/H]. This certainly appears to be the case
for stars with [Fe/H] $\geq -$1, but more metal-poor stars
present an interesting issue, especially those with [Fe/H] $\leq
-$2.5.

Edvardsson et al. (1993) showed that in their sample of disk
stars ([Fe/H] $> -$0.8), relative abundances ([$s$/Fe]) were
solar to within about 0.1 dex. Studies of more metal-poor stars
were reported by Zhao \& Magain (1991) for Y and Zr, and Gratton
\& Sneden (1994) for a sample of ls and hs elements. In the range
[Fe/H] of $-$1 to $-$2, there do appear to be slight departures
from [$s$/Fe] = 0. Gratton \& Sneden report mean values from a
combination of their data with that of Zhao \& Magain's
measurements: $\langle$[$s$/Fe]$\rangle$ = +0.07 (Sr), $-$0.17
(Y), +0.20 (Zr) for the light-$s$ elements, with uncertainties
($\sigma$) of individual values of about 0.1 dex. For heavy-$s$,
Gratton \& Sneden from their own results give
$\langle$[$s$/Fe]$\rangle$ = $-$0.08 (Ba), $-$0.11 (La), +0.04
(Nd), +0.14 (Sm), also with an uncertainty of about 0.1 dex.
Standard errors of these means are about 0.04 dex. These
uncertainties are so much smaller than those associated with the
measurement of abundances in intrinsic and extrinsic $s$-enriched
stars that we ignore the small corrections that might be applied
to the measured [$s$/Fe] to account for departures from [$s$/Fe]
= 0 in the star's original material. One
exception to this is made when attempting to reproduce the
detailed distribution of abundances, including several elements,
as in the three cases discussed in \S~4.

Metal-poor stars  with [Fe/H] $< -$2.5 show an extraordinary
range in heavy element abundances (McWilliam et al. 1995, 1996;
Ryan, Norris, \& Beers 1996; McWilliam 1997, 1998). At a given
[Fe/H], ratios such as [Sr/Fe] and [Ba/Fe] range over 2 dex or
more, from almost $-$2 to +1. McWilliam (1998) showed that
[Ba/Eu] has essentially a unique value despite the enormous range
in their abundances (relative to Fe) with that value being the
solar $r$-process value. Most of the stars have a  solar
$r$-process mix of heavy elements, a mix inherited from their
natal clouds that were contaminated by ejecta from perhaps one or
two Type II SNe. In light of the large range in the heavy element
to iron abundances, a signature of severe $s$-processing may not
always be evident; for example, addition of substantial amounts of
$s$-processed material to a star having an initial [Ba/Fe] $\sim$
$-$2 could still leave the star with a [Ba/Fe] less than the
maximum shown by unevolved stars. With certain assumptions (i.e.,
Eu is purely a $r$-process product and the $r$-process nuclides
are present in solar proportions), it is possible to attempt a
resolution of the elemental abundances into $s$- and $r$-process
contributions (Burris et al. 2000). These attempts may however be
hampered by the recognized bimodal distribution of $r$-elements,
the lighter species (with atomic mass weight below 130-140) being
produced by a different and rarer subclass of Type II supernovae
than the heavier nuclei (Wasserburg \& Qian 2000). In this paper,
we restrict discussion to stars with a [Ba/Fe] well outside the
range spanned by the majority of the very metal-poor stars. We
further eliminate stars for which severe enrichment by the
$r$-process is the more plausible explanation. A few stars do
appear to be highly enriched in $s$-process products. These are
either stars on the main sequence or low luminosity red giants,
i.e. they are either binary (e.g., metal-deficient Ba stars) or
were born out of gas enriched with ejecta from AGB stars. In the
latter case, the meaning of `extrinsic' must be extended and
should not be assumed as always synonymous with mass transfer
binary. The only possibility that such stars are intrinsically
$s$-enriched is that a neutron source is tapped prior to the AGB
phase. The He-core flash is a conceivable suspect, but models
have not so far confirmed this idea. The enormous range in
[$s$/Fe] at a given [Fe/H] is a complicating factor in the
identification and interpretation of $s$-enriched stars, whether
intrinsic or extrinsic.

\subsection{$s$-Enriched Stars of the Galactic Disk and Halo}

Galactic-disk $s$-enriched stars, here [Fe/H] $> -$1, were
considered in Paper II. They include the intrinsic MS and S stars
with technetium, and the cool carbon stars. Extrinsic stars
include MS/S stars with no Tc, the Ba\,{\sc ii} giants, and the
CH subgiants. Here, we collate the observational results of the
highest quality. With two exceptions, we limit discussion to
analyses based on spectra recorded on Reticons or CCDs. The two
exceptions are Smith's (1984) survey, and the use of Tech's (1971)
results for $\zeta$ Cap by Smith \& Lambert (1984). Among the now
excluded stars are the cool carbon stars for which heavy element
abundances are uncertain owing to severe molecular line
blanketing.  For halo stars, here [Fe/H] $< -$1, we gather
results from the literature. The observed values for the
parameters [ls/Fe], [hs/Fe], [hs/ls] are listed in Table 2,
together with references to the original observations. This table
forms the set of observational constraints to theoretical
predictions, extending over the whole metallicity range in the
Galaxy. We have attempted to separate the stars into the
categories extrinsic and intrinsic.

Brief remarks on the stars in Table 2 follow: the sample is
certainly not complete, but we hope it is at least as homogeneous
as possible, and takes into account the most representative
$s$-enhanced sources in the Galaxy. In our comments below we begin
with several classes of what are considered to be intrinsic stars:

{\bf MS/S stars with Tc, Table 2a.} For a sample of intrinsic
stars of disk metallicity, we select the MS and S stars that
exhibit Tc lines in their spectra. Presence of Tc implies that
these stars are experiencing TDUs, i.e., are intrinsically $s$
enriched. Smith \& Lambert (1985, 1986, hereafter SL85 and SL86)
and SL90 analysed 11 MS/S stars and a sample of normal M stars
differentially  with respect to the late-K giant $\alpha$ Tau.
The result that the normal M stars gave  the indices [ls/Fe],
[hs/Fe], and [hs/ls] very close to zero in the mean, with a small
scatter, implies that the abundance obtained for the MS/S stars
are reliable. In the stars from SL90, only Nd was measured among
the `hs' elements, and its abundance is reported in Table 2a. In
the other stars from SL85, SL86, Ba and Nd were used to define
the `hs' parameter and we follow this choice.

{\bf SC Stars, Table 2a.} Abia \& Wallerstein (1998) analysed
seven SC stars and four S stars. Their abundances of the heavy
elements are consistent from one star to the next, but are
somewhat unusual compared to other intrinsic and extrinsic stars.
For [ls/Fe] and [hs/Fe] we find, using Y and Zr, and La and Nd
(where measured) means from six stars of +1.1 and +1.2 for an
average [Fe/H] of 0.1 dex. More typical values in S giants are
around 0.6 dex for these quantities. The differences for SC stars
might in principle be due to a more advanced evolutionary stage,
in which the C enrichment has (almost) reached the condition C/O
= 1. As expected, the mean [hs/ls] index is less affected and the
average value of 0.0 is more consistent with that from other AGB
stars. We must however mention that the spectra of these SC/S
stars are very complex, and rich in CN lines in the red regions
analysed by Abia \& Wallerstein. The difficulty of identifying and
measuring weak atomic lines on these spectra of intermediate
resolution is another possible explanation for the apparently
anomalous composition deduced. As an illustration of the impact
of line blending, we note that few Fe\,{\sc i} lines were
measured: the sample ran from one line in BH~Cru and GP~Ori to a
maximum of 11 for CY Cyg. Very few of the Fe lines were weak. Our
conjecture is that the Fe abundances might have been
overestimated, which would account for the fact that 5 of the
stars had [Fe/H] $\geq$ 0.2. The fact that reference stars -  the
M2.5II-III standard $\beta$~Peg and the K5III star $\theta$~UMi -
gave credible results provides partial confirmation of the
analytical techniques, but the spectra of these stars are not
severely impacted by molecular line blanketing. A recent analysis
of C(N) stars by Abia et al. (2001) further discusses the
uncertainties in measuring C-rich spectra and partially
reconsiders the SC data, suggesting the possibility of
unidentified blends affecting the quantitative results.
Due to this, we give lower weight to the results for these SC/S
stars and represent them collectively, through their mean
abundances (see Table 2a and Figures 7, 8, and 9).

{\bf Post-AGB C-rich Stars, Table 2a.} The label `post-AGB' is
applied to a variety of luminous warm stars. Here, we pick out
those that are rich in carbon with most identified from IRAS
surveys as showing a 21$\mu$m broad emission feature. Six stars
were analysed by Van Winckel \& Reyniers (2000), two by Reddy,
Bakker, \& Hrivnak (1999), and one by Za\v{c}s, Klochkova, \&
Panchuk (1995). These span the metallicity range [Fe/H] = $-$0.3
to $-$1.0 and, hence, are not Galactic halo stars by our simple
classification. All are assumed to be intrinsic stars on account
of their C and $s$-enrichment, attributed to TDUs when on the
AGB. Assuming that departure from the AGB driven by envelope
reduction through a stellar wind occurred after the envelope has
attained its asymptotic mix of $s$ elements (i.e., [hs/ls] is
then a constant independent of the number of TDUs, but [$s$/Fe]
continues to increase), the $s$-process enrichment traces the
terminal ratio [hs/ls] as a function of [Fe/H], and, perhaps,
also the number of TDUs as a function of [Fe/H]  before the star
leaves the AGB. We comment further on these relations later.

{\bf Intrinsic metal-poor carbon stars, Table 2b.} A series of
analyses on such stars have been reported by Kipper and
colleagues. Kipper et al. (1996) discuss five stars with [Fe/H]
between $-$0.7 and $-$1.15. Kipper \& J$\o$rgensen (1994)
analysed a much more metal-poor star ([Fe/H] = $-$2.5). Two
earlier analyses (Kipper \& Kipper 1990; Kipper 1992), not
included in our Tables because they were based on photographic
spectra, were also of very metal-poor stars, [Fe/H] = $-$2.8 for
HD~13826 (V~Ari), and $-$2.9 for HD~112869 (TT~CVn).
Interpretation of these results must bear in mind that molecular
line blanketing in the spectra of these cool giants is severe
and, even in the case of the most metal-poor stars, limits the
accuracy of the elemental abundances: `metal abundances have
quite large errors amounting to 0.5 dex due to extremely heavy
blending of most metal lines by molecular lines' (Kipper \&
J$\o$rgensen 1994) but `in some cases' (say Fe) may be 0.2 dex.
The nature of blending-based errors is such  that they cannot
always be expected to be reduced in forming ratios such as
[hs/ls]. As in the case of SC stars, we assign therefore lower
weight to such results. The status of these stars is a little
uncertain. Despite their temperatures, somewhat in excess of those
for typical AGB stars, `Possibly all cool halo carbon stars may
have formed as intrinsic carbon stars' (Kipper et al. 1996). We
therefore suppose them to be intrinsic, though this assumption
should be better checked in the future. Being old objects, they
should be of rather low mass, but TDU becomes more efficient for
low metallicities, so that the C-star phenomenon might extend to
lower masses for halo stars. Indeed it is known that the number
ratio between C stars and M giants increases at low metallicity.
They are rich in $^{13}$C, like the rather common CJ stars: this
can however be understood at low mass through the operation of
Cool Bottom Processes (CBP: see Wasserburg, Boothroyd, \& Sackmann
1995). Notice that, according to Kipper et al., the stars in Table
2b have to be considered as belonging to the halo, even if their
metallicity does not always obey the rule [Fe/H] $\le -$1. As the
highest metallicities of the Galactic halo and the lowest of the
Galactic disk partially overlap, this rough discrimination is in
this case insufficient. The same comment apply to HD 209621 (see
later Table 2d).

{\bf Extrinsic stars in the Galactic disk, Table 2c.} We include
here MS and S stars without Tc in their spectra, which are
supposed to be evolved versions of barium stars. Indeed,
abundance analyses by SL85, SL86, SL90. show that they have
$s$-enrichments like those of Ba\,{\sc ii} K-type giants. The
same rules discussed for MS/S stars with Tc on the selection of
ls and hs abundances apply here. We also consider the CH
subgiants analysed by Smith et al. (1993), and Ba\,{\sc ii}
giants from several sources (Smith 1984; Smith \& Lambert 1984;
Tomkin \& Lambert 1983, 1986; Kov\'{a}cs 1985; Smith \& Suntzeff
1987). For CH subgiants, we include Ba and Nd in the `hs' index,
as done by Smith, Coleman, \& Lambert (1993). Due to our already
mentioned choices on the type of observations to include, we omit
extrinsic stars from Luck \& Bond (1984, 1985, 1991a,b), because
abundances were derived from photographic spectra. As shown
elsewhere (see e.g. Busso \& Gallino 1997), those abundances can
be in general well explained by AGB models; they however give
constraints qualitatively similar to those already provided by
our more homogeneous set of data. In the class of extrinsic
Galactic disk stars we may include also the K giant
BD~+75$^\circ$348 (Za\v{c}s, Schmidt, \& Schuster 2000), which
has [Fe/H] = $-$0.8 and $s$ enrichment similar to those of an
extreme Ba\,{\sc ii} star. One supposes it is a spectroscopic
binary but radial velocity measurements are presently lacking.

{\bf Extrinsic Galactic halo AGB stars, Table 2d}

In the group of Galactic halo extrinsic stars we include the
objects described below.

{\bf CH giant stars.} A sample of these stars, the metal-poor or
halo counterpart of the classical Ba\,{\sc ii} giants, were
analysed by Vanture (1992). Abundances of heavy elements were
reported for seven stars with [Fe/H] from $-$0.9 to $-$1.7. A
revision for HD~187861 was also suggested by Vanture (2000,
private communication). Their status as `extrinsic' is confirmed
by their designation as spectroscopic binaries (McClure 1983,
1984). Owing to the larger uncertainty in abundance determinations
in these stars (in particular HD 209621 and HD 221959), we assign
lower weight to these results.

{\bf Yellow symbiotic stars.} Three such stars have been analysed
recently: AG Dra (Smith et al. 1996), BD~$-$21$^\circ$3873 (Smith
et al. 1997), and He2$-$467 (Pereira, Smith, \& Cunha 1998).
`Yellow' denotes that the giant has a spectral type of G-K and,
hence, its spectrum is not cluttered with the TiO lines found in
redder symbiotic stars. It is fair to presume that the giant has
accreted material from its companion, which is now a white dwarf
but earlier was  a mass-losing AGB star.

{\bf HD~196944.} Za\v{c}s, Nissen, \& Schuster (1998), who
analysed this star, considered it `unlikely' that it is a post-AGB
star and place it as a peculiar AGB star or a CH star (i.e.,
extrinsic).

{\bf  LP 625-44,  LP 706-7, and CS 22898-027.} The first is a
C-rich star with [Fe/H] = $-$2.7, analysed by Norris et al. (1997)
and Aoki et al. (2000). With [Ba/Fe] = +2.7, it is about 2 dex
more Ba-rich than the most Ba-rich `normal' metal-poor stars
(McWilliam et al. 1995), and the ratio [Ba/Eu] = 0.8 implies
$s$-process enrichment rather than   $r$-process contamination
(for which [Ba/Eu] = -0.8 in the solar case, which seems to be
representative of the heaviest elements  in metal-poor
$r$-process-dominated stars).  Aoki et al. consider, partly on the
grounds of radial velocity variations, the star to be the `result
of mass transfer in a binary system from a previous AGB
companion' (i.e., LP 625-44 is an extrinsic star.).
 Abundances are taken from Aoki et al. It is
clear that abundances from Ba to Pb follow the pattern expected of
$s$-processing in a metal-poor AGB star of low mass. The low
abundance of Europium, a traditional tracer of the $r$-process, is
well fit by the predictions of the marginal $s$-process
contribution, which amounts to only 5\% of solar Eu. The C-rich
star LP 706-7 (Norris et al. 1997) with [Ba/Fe] = 2.0 is similar
to LP 625-44, but slightly less evolved. It might be considered
extrinsic, but proof in the form of radial velocity variations is
presently lacking. CS 22898-027 (McWilliam et al. 1995; McWilliam
1998) is similarly C-rich and $s$-process enriched.

In contrast to this trio, there are stars, even C-rich stars,
whose pattern of heavy element overabundances is that of the
$r$-process. The best known examples are CS 22892-052 (Sneden et
al. 1996), HD\,115444 (Westin et al. 2000), and CS 31082-001
(Cayrel et al. 2001; Hill et al. 2001), but other similar objects
exist (see e.g. Hill et al. 2000; Barbuy et al. 1997). Relative
abundances of the heavy elements Ba to Pb match well the solar
$r$-process abundances; the low [Ba/Eu] ratio is that expected of
a solar-like $r$-process. Any possible $s$-process contribution
to the abundances is swamped by the dominant $r$-process
contribution. It should be noted that in very metal-poor stars,
while the relative abundances of heavy elements are remarkably
similar to those in the solar $r$-process, the ratio of the
heavy-$r$ to light-$r$ elements is variable from star to star
(Qian, Vogel, \& Wasserburg 1998; Burris et al. 2000, Qian \&
Wasserburg 2001). As mentioned, this has been interpreted in the
framework of a bimodal $r$-process, coming from supernovae of
different frequency (Wasserburg \& Qian 2000, Sneden et al. 2000).

A pair of stars showing evidence of carbon (and nitrogen)
enrichment with heavy element abundances intermediate between
pure solar $r$-process and an $s$-process was analysed  by Hill
et al. (2000). That the ratio [Ba/Eu] $\simeq$ 0.0 is
intermediate between that of the $r$-process dominated CS
22892-052 ($\simeq$ $-$0.8) and that of the $s$-process dominated
LP  624-44 ($\simeq$ +0.8) implies the atmospheres of the pair
are a blend of $r$- and $s$-processed material. Unfortunately,
the mentioned fact that there is not a unique set of $r$-process
abundances for light and heavy elements bedevils attempts to
resolve the abundances of Hill et al.'s stars into their $r$- and
$s$-process components. In particular, the star-to-star variation
in the ratio of the light $r$- to heavy $r$-elements necessarily
compromises the extracted ratio of the light $s$ to heavy $s$
elements. Until the star-to-star variation is better understood,
selection of $s$-process enriched very metal-poor stars is
limited to those stars where the $s$-process is obviously
dominant. Although the present sample is extremely small, there
is a tantalising hint that the $r$-process enriched stars like CS
22895-052 are giants, and the $s$-process enriched stars like LP
625-44 are dwarfs or subgiants.

{\bf N-rich subdwarfs: HD~25329 and HD~74000.} The rare class of
N-rich subdwarfs was discovered by Bessell \& Norris (1982).
Beveridge \& Sneden (1994) analysed  two of these stars. For HD~
25329, they determined the abundances of 9 heavy elements but for
HD~74000 just 3 heavy elements were measured. These stars have
metallicities of $-$1.8 and $-$2.1, respectively. An $s$-process
enrichment is indicated for HD~25329; too few elemental
abundances were given for HD~74000 to make a similar assertion.
The heavy elements, as well as the N, Na, and Al enrichments were
attributed to synthesis in an AGB star but lacking direct
evidence for a companion, the possibility of contamination of the
natal clouds was left open.

\subsection{Observation and Theory}

Estimates of [ls/Fe], [hs/Fe], and [hs/ls] were made for the
stars in the previous section. Our estimates may differ from
published values because we elect to consider a different mix of
elements. The differences are slight and the overall conclusions
drawn from the assembled data are unaffected by whether the
original or our choices of heavy elements are adopted. We adopt
in general the published estimates of [Fe/H]: exceptions are MS/S
giants, where we make use of the more reliable [M/H] values, `M'
indicating an average from Ti to Fe. Using this average changes
only slightly the values of [ls/Fe] and [hs/Fe]. Here, we compare
the $s$-process indices against theoretical predictions.

Run of [hs/ls] versus [Fe/H] is shown in Figure 7, where filled
and unfilled symbols refer to intrinsic and extrinsic stars,
respectively. The two N-rich subdwarfs are shown by asterisks.
Theoretical predictions for ST/1.5 (bold line) provide an
approximate average fit to the data. For a comparison, two more
lines, representing the envelope of our maximal and minimal
predictions for [hs/ls] (as deduced from Figure 3a) are shown as
dashed and dotted lines, respectively. This makes clear that most
observed data are explained, within the errors, by model curves.
The most remarkable outliers in Figure 7 are four C-rich objects
with [Fe/H] $\le -$2 (one probably intrinsic, three probably
extrinsic). They are CS~22898-027, LP~625-44, and LP~706-7 and
HD~187216. The first has been already commented upon: absence of Y
and Zr makes its data rather uncertain in our picture. The last
was recognized as a puzzle difficult to accomodate in any
explanation based on LMS evolution/nucleosynthesis already by
Kipper \& J$\o$rgensen (1994), but has a complex spectrum and
larger-than-average error bars should be attributed to its
abundances. LP~625$-$44 was recently analysed by Ryan et al.
(2001), who found the distribution of $s$-elements from Ba to Pb
in remarkable agreement with nucleosynthesis predictions from very
low mass stars ($M \le$ 1.3 \ms). The same may hold for
LP~706$-$9, but spectra of higher resolution are needed. A very
low initial mass, in connection with possible large
inhomogeneities in the parent ISM at low metallicities, might be
invoked for explaining the whole set of abundances in these
extreme halo C-rich objects. We have problems in the fits also
with IRAS 22272+5435, a post-AGB supergiant from  Za\v{c}s,
Klochkova, \& Panchuk (1995). However, this star has a large
uncertainty in its metal abundances (e.g., Fe is deficient, the
other elements of the Fe group are overabundant, the $s$-element
abundances have uncertainties up to 0.5 dex) so that we must
attribute a lower weight to it.

Conversion of a normal star to an extrinsic star by mass transfer
from an AGB companion is not expected to change the [hs/ls]
index. This expectation assumes that there is no
differentiation/fractionation during the transfer of mass.
Certainly, if dust and gas in the stellar wind are transferred
and accreted differently, one might expect a difference owing to
the selective condensation of elements into and onto grains. A
cautious interpretation of Figure 7 is that intrinsic and
extrinsic stars follow similar trends of [hs/ls] versus [Fe/H].
There is a suspicion voiced first by Reddy et al. (1999) and Van
Winckel \& Reyniers (2000) that the intrinsic post-AGB C-rich
stars have a lower [hs/ls] than the Ba\,{\sc ii} stars and the CH
giants. The two N-rich subdwarfs stand apart from the remainder
of the sample, and are also distinguished by quite mild $s$
enrichments, explained only by models with the lowest $^{13}$C
concentration in the pocket.  In what follows, we consider them
separately from the majority of the sample.

Indices [ls/Fe] and [hs/Fe] are expected to be different for
intrinsic and extrinsic stars with a likelihood that, if the mass
transfer is not selective, extrinsic stars could show lower
values than intrinsic stars of the same metallicity. Remarkably,
the run (Figure 8) of [ls/Fe] versus [Fe/H] follows  the
theoretical prediction for ST/1.5 for intrinsic AGBs, while
extrinsic ones lie on average below the curves, as expected. If
the predictions were reduced by only about 0.2 dex, the fit would
be even more satisfactory, especially for Galactic disk stars.
This can easily be achieved, when one considers that model curves
refer to the final AGB situation, while several intrinsic AGBs
(MS, S stars) are certainly in a previous stage, and therefore
have not yet reached the terminal values of their $s$
enhancements (see later \S~3.4, and also Abia et al. 2001). In
light of Figure 3, a better average trend can alternatively be
obtained by choosing a prediction between the ST/1.5 and ST/3
values. This possibility of accounting for the observed data by
varying either the efficiency of nucleosynthesis or that of
mixing was already encountered in Paper I, where we showed as
sometimes only consideration of the whole distribution of $s$
elements (if available) can solve this ambiguity. Observed and
predicted [hs/Fe] indices are shown in Figure 9. Disk stars,
especially for [Fe/H] $> -$0.5, fit the prediction for ST/1.5
reasonably well, as above. There is a hint that observations for
the intrinsic stars fall below the ST/1.5 model curve around
[Fe/H] $\simeq -$ 1, and the same duplicity of solutions
mentioned for the `ls' abundances applies.

In short, the collated observations of $s$-enrichment in intrinsic
and extrinsic stars are, as a whole, rather well accounted for by
the models of $s$-processing and TDUs in AGB stars. In
particular, predictions for the ST/1.5 case provide, at all
metallicities, a sort of average trend that nicely compares with
observations, when the effects of mixing are taken into account.
Our analysis considered a mass of 1.5~$M_\odot$ for the AGB star.
In general this is not a critical choice; Figures 3 and 4 show a
weak dependence of the predictions on initial stellar mass, as
long as we consider low mass AGB stars. However, some exceptions
exist. Four examples in Figure 7 have been mentioned. Three
additional cases are HR~774, BD~+75$^\circ$348, and
IRAS~07134+1005. We find a solution with 3 \msb models in the
first two cases, and we need a mass lower than 1.5~\msb in the
last one (see \S~4, also for the peculiar case of HD~25329).
Information on the initial mass of the AGB star is provided by
detailed comparisons of observed abundances with theoretical
predictions for [hs/ls], [ls/Fe], and [hs/Fe], as deduced by
Figures 3a,b,c for $M$ = 1.5 \msb and by Figures 4a,b,c for  $M$
= 3 \ms, respectively. Indeed, the values of these parameters
provided by the models are rather sensitive to the initial mass
(and to the metallicity) of the model star adopted.

\subsection{\bf A closer look at the role of mixing.}

Concerning the two parameters (efficiency of mixing and of
nucleosynthesis) that affect the $s$-process distributions of
Figures 7 to 9, we actually have tools to discriminate their
relative role thanks to the fact that models of the envelope
abundances were computed pulse after pulse. An example of this is
shown in Figure 10, which is the equivalent of Figure 2 of Paper I
and Figure 6 of Paper II, but is computed with the new models: it
is built for Galactic-disk metallicity models of 1.5 \msb and
contains only data points for Galactic-disk stars. The continuous
lines show model envelope enrichments up to our last TDU episode,
for different $s$-process efficiencies, as measured by the value
of the number ratio between \ctb and iron in the pocket. In order
to understand the importance of gradual mixing, the dashed lines
connect the points representing the 4th and 8th TDU episode. The
figure shows clearly how many stars have abundances that are best
explained by situations less extreme than those of Figures 3 and
4, which represent the termination points in the distributions of
Figure 10. In particular, several MS/S stars with technetium in
their spectra find a solution only with a limited number of TDU
episodes, as expected by their belonging to a class intermediate
between unprocessed M giants and C stars. A complementary view is
provided by Figures 11 and 12, where the curves (again from the
1.5 \msb case) show model envelope compositions at different
metallicities (similarly to Figure 3) in the plane [s/Fe] versus
[hs/ls]. Here `s' means the average of both `hs' and `ls'
abundances and gives therefore integrated information on the
$s$-element enrichment. The high non-linearity of the models as a
function of the iron content of the star can be inferred by the
indication on the plot of where the high and low metallicities
lie.  The loops correspond to the maxima in Figure 3. While Figure
11 represents the situation at the end of the AGB phase, Figure 12
is constructed with the predictions after only 4 pulses. This last
case shows the composition of a material which is highly diluted,
i.e. contains a small $s$-process enrichment from the He
intershell. Most data points are bracketed by the predictions of
the two plots, showing again that observations present to us a
variety of mixing efficiencies. In these plots model curves allow
one to appreciate only the total dilution, hence cannot
discriminate between, for example, intrinsic MS stars and
extrinsic Ba stars, though only the former class owes its
dilution to an incomplete dredge up of He intershell material,
while the second includes mass transfer in a binary system.

There are particular and intriguing issues  in the plots from
Figure 7 to Figure 12. Two will be mentioned: the case of the
post-AGB stars and the Ba\,{\sc ii} stars, and the N-rich
subdwarfs. As noted earlier (Reddy et al. 1999; Van Winckel \&
Reyniers 2000), there is a suggestion that the post-AGB intrinsic
C-rich stars have a lower [hs/ls] than the Ba\,{\sc ii} giants.
This difference of about 0.3 dex arises because of roughly equal
and opposite offsets of the two groups in the indices [ls/Fe] and
[hs/Fe]. Given that the atmospheres of the two groups of stars
differ in temperature  and surface gravity, the possibility
exists that these small differences in [ls/Fe] and [hs/Fe]
reflect systematic errors in the LTE  analyses of either or both
sets of stars. It is certainly premature to consider other
explanations, among them selective accretion of elements by the
extrinsic star in a mass transfer process.  Detailed comparisons
of observed and predicted abundances of Ba\,{\sc ii} giants could
reveal clues to the possible difference with post-AGB stars; we
discuss one such case in the next section. A further remarkable
point to be mentioned for post-AGB stars is that our envelope
compositions at the last thermal pulse found with the stellar
code can fit them quite well, despite the noticeable envelope
mass remaining in the models when TDU ceases. This may be a
support to the finding that TDU disappears well before the
envelope is completely eroded by mass loss. The physical reason
has been suggested to lay in a sudden instability of the whole
envelope, driven by a sharp reduction of the gas pressure and by
the establishment of a super-Eddington luminosity (Lattanzio,
Forestini, \& Charbonnel 2000; Sweigart 1998; Wood \& Faulkner
1986). A closer look at the N-rich subdwarf HD~25329 is given in
the next section.

\section{Detailed photospheric compositions: a few examples}

As mentioned in the previous subsection, consideration of just the hs
and ls enhancements may leave unclear the relative importance of the two main
processes controlling the abundances, i.e. the efficiency of neutron
captures and the dilution mechanisms (firstly through TDU, subsequently
through mass transfer, for the binary sources).

Disentangling the two effects is made easier when the detailed
set of observations for individual elements is available,
including the CNO group and other light nuclei. Indeed the
synthesis of these last by the red giant is almost entirely
unrelated to the efficiency of the neutron source. In particular,
due to the rather fixed abundance of \cdb in the He intershell at
the occurrence of TDU ($X$($^{12}$C) $\simeq$ 0.20 $-$ 0.23) the
photospheric abundance ratio between carbon and $s$ elements
helps to set independent constraints on the dilution with the
envelope. One must however take into account the complicacies
introduced by the possible activation of H-burning processes in
the envelope, that can burn $^{12}$C and produce $^{13}$C and
$^{14}$N. These are HBB, occurring in stars of intermediate mass
(Lattanzio \& Forestini 1999; Lattanzio, Forestini, \& Charbonnel
2000) and CBP in stars of mass $M$ $\le$ 2.3\msb (Gilroy 1989;
Gilroy \& Brown 1991; Charbonnel 1995; Wasserburg, Boothroyd, \&
Sackmann 1995; see also discussion in BGW99). In IMS, when HBB is
at play, the ratio \cd/\ctb decreases to its CNO equilibrium
value of $\sim$ 3.5 and remains rather low even during the TP-AGB
phase. Concerning CBP in LMS, it reduces the ratio \cd/\ctb
already during the red giant phase, and its effects depend on the
initial mass. According to observations of giant stars in
Galactic Clusters, this ratio is $\sim$ 12 for $M$ = 1.5 \ms. It
decreases to the CNO equilibrium value for giant stars of $M$
$\le$ 1 \msb in Globular Clusters. During the AGB phase, it is
possible for CBP to continue its operation, thus converting part
of the primary $^{12}$C mixed with the envelope by TDU into
$^{14}$N (Wasserburg, Boothroyd, \& Sackmann 1995).

When our analysis is extended toward low metallicities, the task
of reproducing carefully the abundance distributions of
individual sources by operating such a dilution is complicated by
the need to consider the possibly highly non-solar initial
composition, characterized by a large scatter in the initial
abundances of heavy nuclei. Here it is worth showing some
examples, in order to demonstrate that, despite this, new pieces
of information can be achieved on the mass and evolutionary
status of the target stars.

We show therefore, in Figure 13, 14, 15,  three examples of a thorough
comparison between nucleosynthesis predictions and spectroscopic
observations. For the sake of these examples, the initial
abundances of the stellar envelopes have been scaled to the
appropriate metallicity, following the already mentioned average
trends of heavy element abundances in the Galaxy, published by
Gratton \& Sneden (1994).

The cases shown were chosen because of their heterogeneity
(representing high, intermediate and low metallicities; including
a Ba star (extrinsic AGB), a post-AGB supergiant and a suspected
extrinsic dwarf; and belonging to the giant, supergiant, and
dwarf luminosity classes, respectively). Their characteristics and
detailed fits are discussed below.

\subsection{HR~774: a Ba II giant star in the Galactic disk}

After the pioneering work by Warner (1965), a large number of
studies considered the heavy element abundances in classical Ba
II giants.  HR~774 being very bright ($m_v$ = 5.96 according to
Smith 1984), has been the object of several spectroscopic
observations (Tomkin \& Lambert 1979, 1983; Smith 1984). Smith
(1984) emphasised the relatively low $^{12}$C/$^{13}$C ratio, a
feature  noticed in extrinsic AGB stars (in CH stars) by
Wallerstein \& Greenstein (1964). The set of observations adopted
here is the one compiled in Paper II (Table 2) through averages
of the various original observations. In particular we accept a
metallicity [Fe/H] = $-$0.3~$\pm$~0.2. The data are modeled by
generating the expected photospheric composition by dredging up
carbon- and \s-process-rich material from the He intershell to
the envelope, according to the TDU efficiency of the stellar
code, pulse after pulse. As shown in Figure 13, our best fit to
the abundances comes from a binary system having a primary
component of initially 3 \ms, with a rather high efficiency in
$s$ processing (ST$\times$2), and a low mass secondary. In this
example, the same results are obtained with two different mass
transfer approaches. One assumes a single phenomenon of
accretion, at the 20th TP, with a dilution of 0.2 (i.e. one part
of the AGB envelope per 5 parts of the atmosphere of the
secondary component). The second assumes a continuous process, up
to the 20th pulse, and yields a dilution of 0.4. Due to the
uncertainties on mass loss rates and our poor knowledge of how
and when mass transfer phenomena occur, we do not claim that
these are the only, or even the best solutions. They, however,
indicate that a moderately massive primary component is necessary
here. In this way we reach an agreement also for the carbon
abundance, (values [C/Fe] in the range 0.5$-$0.7 can be obtained,
depending on the assumptions made, to be compared with the
observation 0.7~$\pm$~0.1). We found this datum difficult to
reproduce otherwise.

\subsection{IRAS~07134+1005: a post-AGB supergiant}

This star belongs to the group of supergiants showing the 21
$\mu$m feature in their mid-infrared spectra, which is suspected
to derive from C-rich compounds (Kwok, Volk, \& Hrivnak 1989).
Observations of molecular bands from CO, CN, HCN and C$_2$ have
been reported by various authors (Bakker \& Lambert 1998a,b;
Bujarrabal, Alcolea, \& Planesas 1992), so that both the C/O ratio
and the $^{12}$C/$^{13}$C ratio are known. Optical spectroscopic
observations of high precision were recently presented by Van
Winckel \& Reyniers (2000). They found it impossible to reproduce
the measured $s$-element abundances with the parameterized
calculations by Malaney (1987). The more complex $s$-process
distributions of stellar models can instead perform this task,
using cases of the appropriate metallicity ([Fe/H] = $-$1). In
the exercise shown in Figure 14, we found a good fit to the data
with a 1.5 \msb model with a relatively low efficiency in $s$
processing (ST/3). With our choice for mass loss the fit requires
10 pulses; at this phase also the carbon and nitrogen enrichment
are compatible with observations ([C/Fe] = 1.3, [N/Fe] = 0.9,
where observed values are 1.1~$\pm$~0.1 and 0.85~$\pm$~0.15). The
suggested number of pulses is highly model dependent. Anyway, it
indicates that the parent star was probably of initial mass lower
than 1.5 \ms, therefore experiencing only a limited number of TDU
episodes.

\subsection{HD~25329: a star from an $s$-enriched cloud or a metal-poor Ba
dwarf?}

HD~25329 is currently classified as a halo dwarf, and is not
particularly C rich, but very N rich. Beveridge \& Sneden (1994)
conclude that the star has a metallicity [Fe/H] = $-$1.8
(possible uncertainty is of the order of $\pm$0.2~dex). These
authors derived C, N, and O abundances as well as abundances of
light, iron-peak, and heavy elements. Gay \& Lambert (2000)
showed that HD~25329 is marked by an apparently high abundance of
the isotopes $^{25}$Mg and $^{26}$Mg relative to $^{24}$Mg. It
shows $^{26}$Mg/$^{24}$Mg $\simeq 0.09$, where this ratio is 0.03
in the normal subdwarf HD~103095 of slightly higher metallicity.

Spectroscopic observations revealing enrichment in heavy elements
were early presented by Peterson (1981a,b), who attributed them
to explosive nucleosynthesis. Subsequently, Beveridge \& Sneden
(1994) obtained high S/N spectra from which they derived the
abundances of  $s$ elements that we analyse here. These authors
suspected that the enhancements of heavy nuclei might originate
in the He-burning shell of an AGB star, but, as they could not
confirm this star to be a spectroscopic binary, suggested that
HD~25329 formed from gas polluted by ejecta from one or more AGB
stars. The higher than expected abundance of $^{25}$Mg and
$^{26}$Mg (relative to the normal abundance of $^{24}$Mg)
suggests that the ejecta came from intermediate mass AGB stars.
Our attempts to fit the abundances of this $s$-enriched dwarf may
confirm this conclusion (see below).

We addressed the problem through two exercises, whose results are
shown in Figure 15. In the first attempt (lower panel), excluding
for the moment Mg isotopes, we succeed in reproducing the
observed $s$-process data within their uncertainties, as well as
the relatively low C abundance, by adopting a model of very low
efficiency in neutron production (ST/12). The fit shown adopts an
initial mass $M$ = 1.5 \ms. The $s$ element abundances must not
grow too much in the AGB primary star, because the observed
enhancements are rather small. This can be obtained, for example,
if the primary component experiences only a limited number of TDU
episodes (less than 10, as assumed in the Figure; this may be
rather typical for these metallicities and also allows one to
limit the carbon enrichment, which in our model is [C/Fe] = 0.5,
against an observation of 0.6 ~$\pm$~0.1). A good fit to the $s$
abundances can be obtained either with a unique mass transfer
episode at the 8th TDU episode (yielding a dilution of 0.33, i.e.
one part of the AGB envelope over three parts of the companion's
atmosphere) or with a continuous addition, lasting from the first
to the eighth TDU episode (and in this case a larger fraction of
the observed material would come from the AGB envelope, yielding
a dilution of 0.54, due to the lower $s$ enrichment in the first
pulses). These details are however very model dependent. In this
example the enhancements of N and $^{25,26}$Mg cannot be
explained, and we would be forced to assume that these anomalies
are inherited from the parent cloud. In a second exercise (upper
panel) we showed that $s$-process abundances might also derive
from a mass transfer episode from an IMS primary of initially 5
\ms. This second exercise needs a noticeable $^{13}$C
concentration in the pocket (for the assumptions in IMS and the
definition of what is the `ST case' there, see e.g. Travaglio et
al. 2001). This solution is however not completely satisfactory
for the light elements. Indeed, to reach the observed abundances,
we have to assume that several pulses were made by the primary
component, to compensate for the large dilution in the massive
AGB envelope. This leads to a C-rich secondary (C/O $>$ 1),
contrary to the observations. We might obviously assume that HBB
burns this extra carbon (Lattanzio \& Forestini 1999), but we did
not apply a proper model to this. Also, $^{26}$Mg becomes too
high, with a ratio $^{26}$Mg/$^{24}$Mg close to 0.3 for an
initially solar Mg isotopic mix.

We therefore prefer the idea advanced by Beveridge \& Sneden
(1994), according to which this source might have formed from a
natal cloud previously contaminated in $s$-process elements by a
few AGBs, including at least one IMS. Unfortunately, our arguments
are not strong enough to exclude completely the alternative
hypothesis that it is a low-metallicity Ba dwarf, born with
somewhat anomalous light element abundances. From Figure 15 we
can however at least get a tool to discriminate the mass of the
contaminating source(s). As shown by the two distributions, the
requirements imposed by the other $s$ elements yield in the two
cases very different predictions for Pb, which is expected to be
very high if IMS were at play here. We leave this tentative
suggestion to future observational tests.

\section{Conclusions}

In this paper we have shown how present AGB modeling can
satisfactorily account for the $s$-element enrichment observed at
the surface of intrinsic and extrinsic AGB stars of different
metallicities, from solar down to extreme halo composition. Agreement
between model predictions and average observed abundances for
heavy (Ba-peak group) and light (Zr-peak group) $s$ elements (hs and
ls) can be obtained by varying over a rather narrow interval the abundance of
$^{13}$C burnt per nucleosynthesis episode.
The almost linear relationship between the neutron-flux sensitive
parameter [hs/ls] and [Fe/H], which roughly holds at Galactic disk
metallicities, breaks down for halo stars. The reason for this is
primarily that at very low [Fe/H] values the number of neutrons
captured by iron seeds and their progeny becomes so large that
the $s$-process path is followed to its end at $^{208}$Pb (and at
a lower extent at $^{209}$Bi). The [hs/ls] parameter grows with
the total $s$ enrichment [s/Fe] in the envelope, with a functional
relationship that appears roughly linear for post-AGB stars (as
noticed by Van Winckel \& Reyniers 2000), but in reality displays
a loop-like trend and is different for AGB and post-AGB stars,
reflecting different dilution factors of the $s$-processed matter
with the envelope. We also showed how the models can account for
the detailed distributions of $s$-elements and for the abundance
of CNO nuclei, when they are available, and how this type of
comparisons can add new pieces of information concerning the
initial mass, the number of thermal pulses occurred and the
dilution efficiency for extrinsic AGB stars.

\acknowledgements

This work was supported by MURST through the grant
COFIN2000$-$Evoluzione Stellare, the U.S. NSF through grants to
DLL (grant AST96-18414) and to VVS (grant AST96-18459), and the
Robert A. Welch Foundation.  We are deeply indebted to O.
Straniero, A. Chieffi and M. Limongi for their longstanding
fundamental collaboration in stellar modeling. We are also
grateful to the referee, Beatriz Barbuy, for her helpful and
constructive comments.

\newpage

\plotone{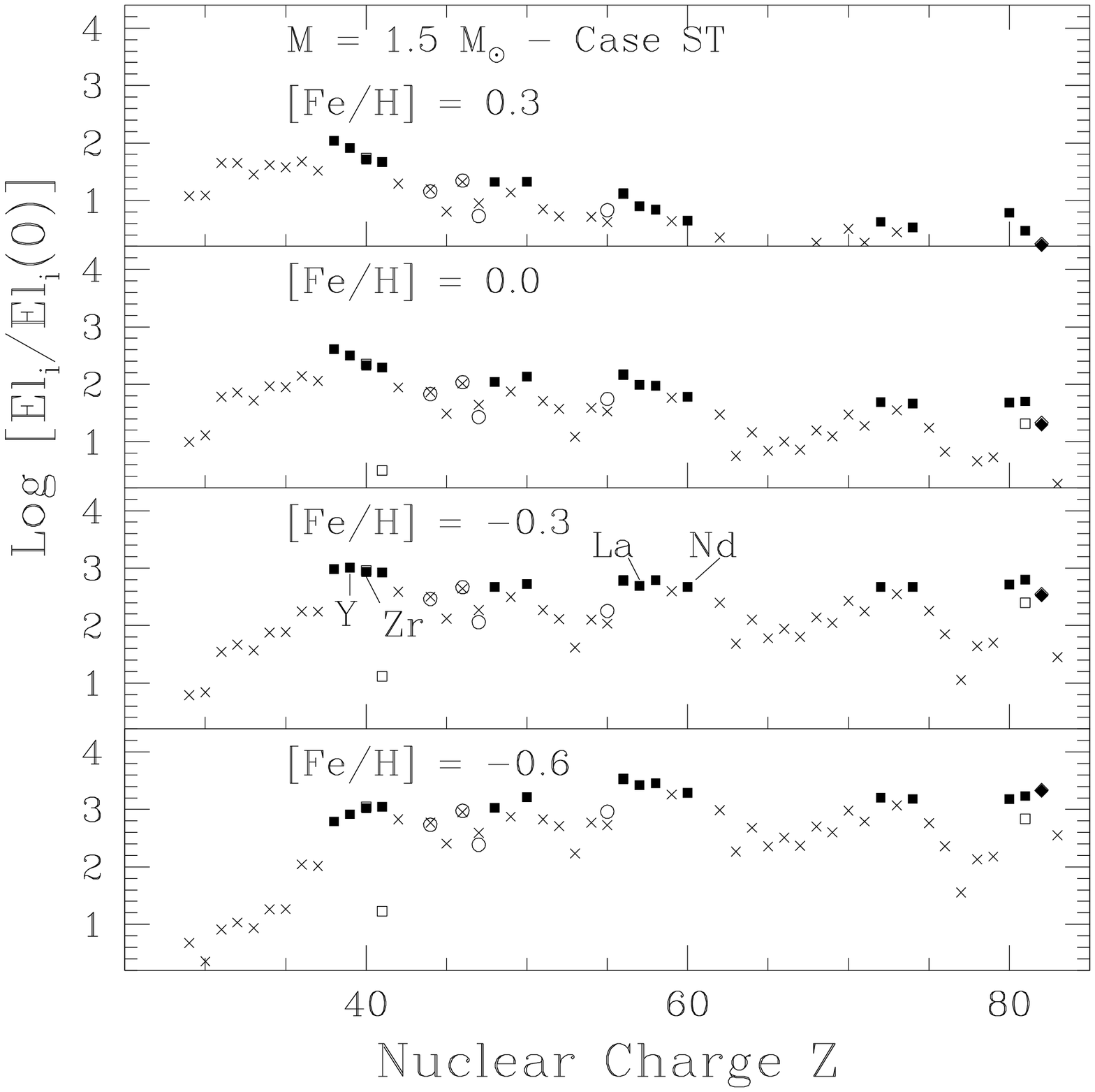}
\figcaption{{\small The distribution of the elements from Cu to Bi
in the He
intershell as a function of the initial metallicity. Elements
mainly produced by the main component of the $s$ process (by at
least 50~\% according to Arlandini et al. 1999) are shown as bold
squares. Pb has a special symbol (bold diamond) because it is
mainly attributed to the strong $s$-process component, deriving
from AGB stars of low metallicity (Travaglio et al. 2001). The
models refer to a 1.5 \msb star, with the choice ST for the \ctb
pocket discussed in the text, and with metallicities from [Fe/H]
= +0.3 down to $-$0.6. For elements affected by the decay of
unstable isotopes whose half-life is longer than a typical
interpulse period, but shorter than 10$^9$ yr, open symbols refer
to abundances before decay.}}

\newpage

\plotone{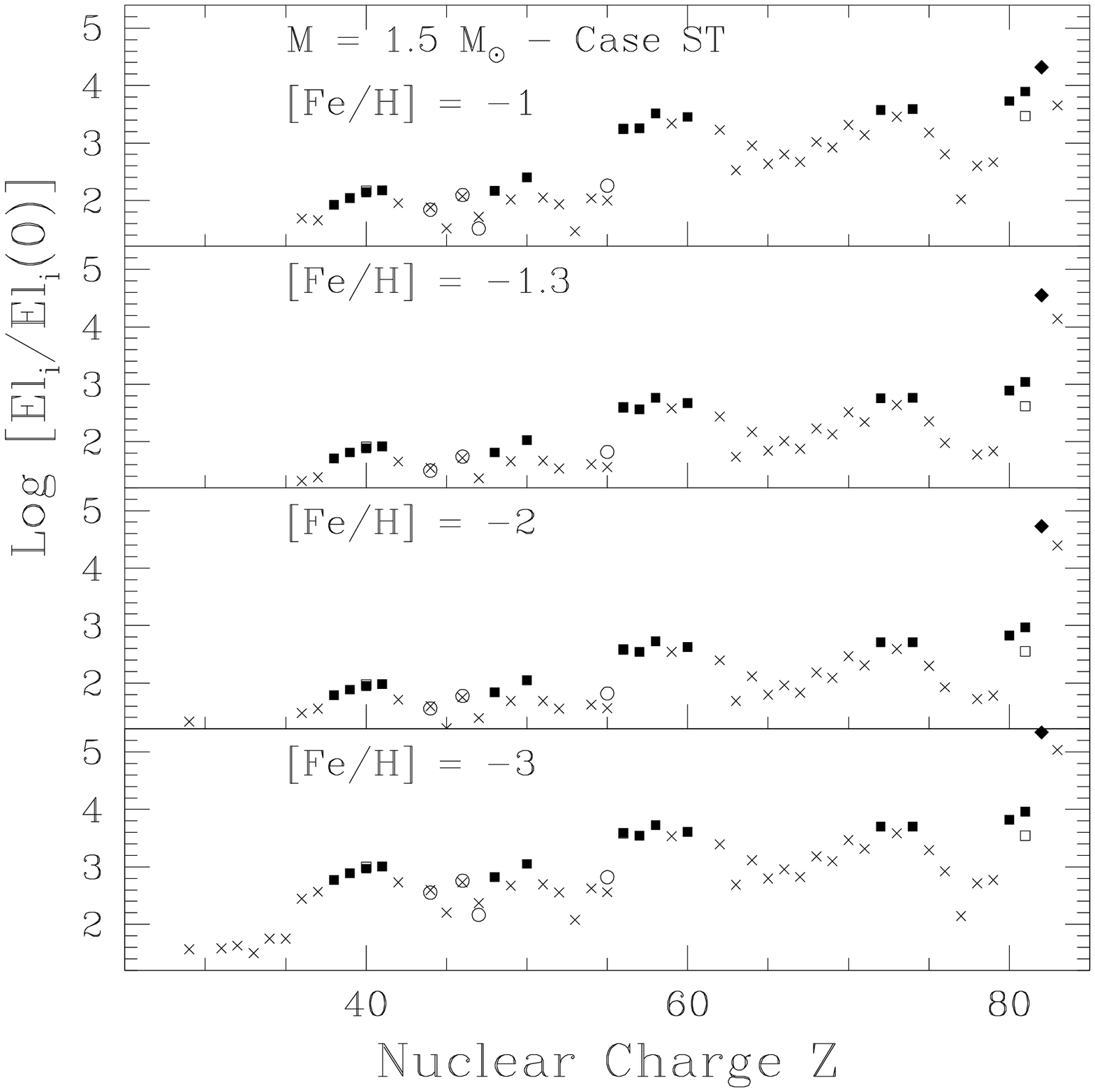}
\figcaption[f2.eps]{Same as Figure 1, but for metallicities from [Fe/H] =
$-$1 down to [Fe/H] = $-$3.}

\newpage

\plotone{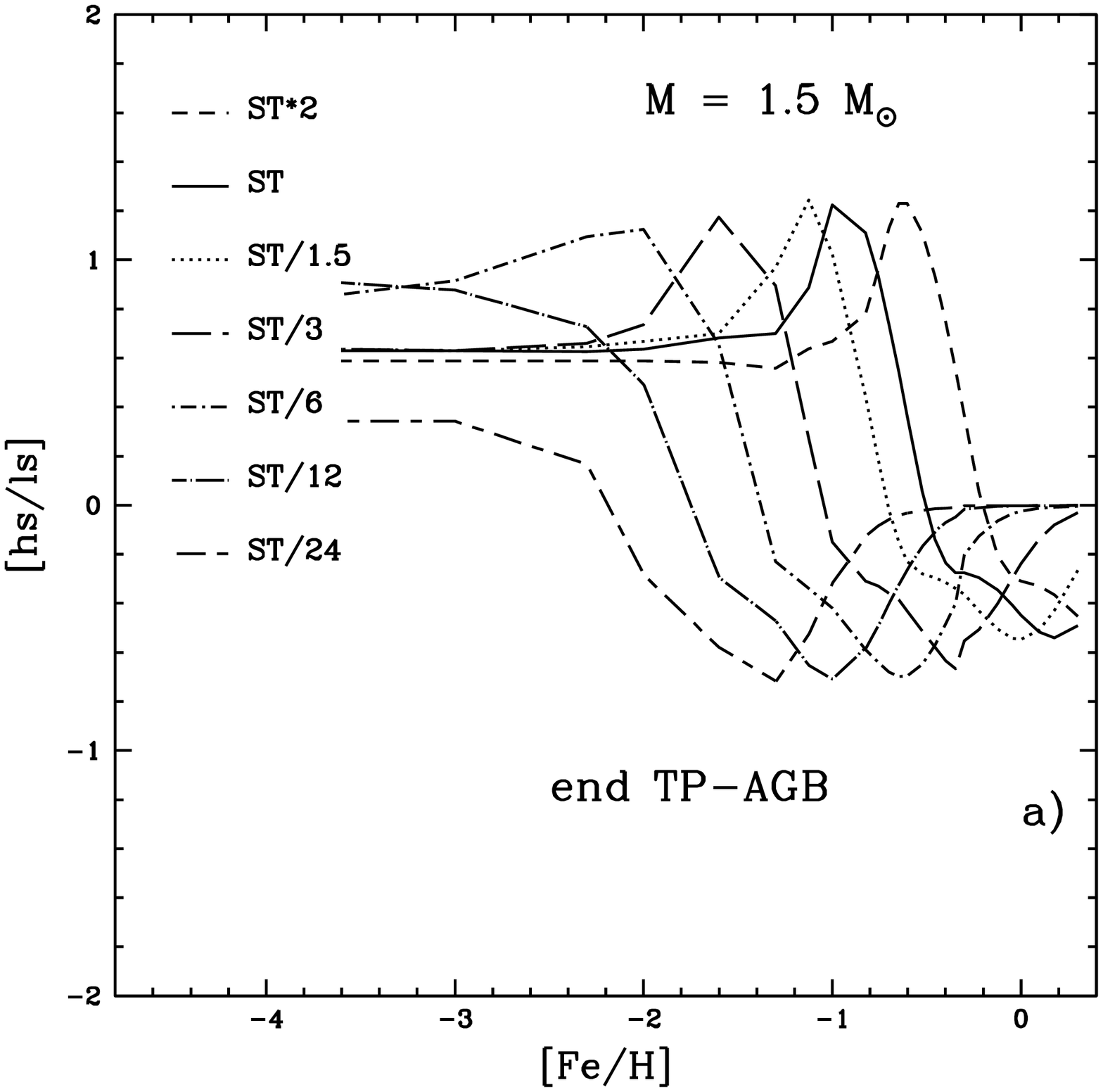}

\newpage

\plotone{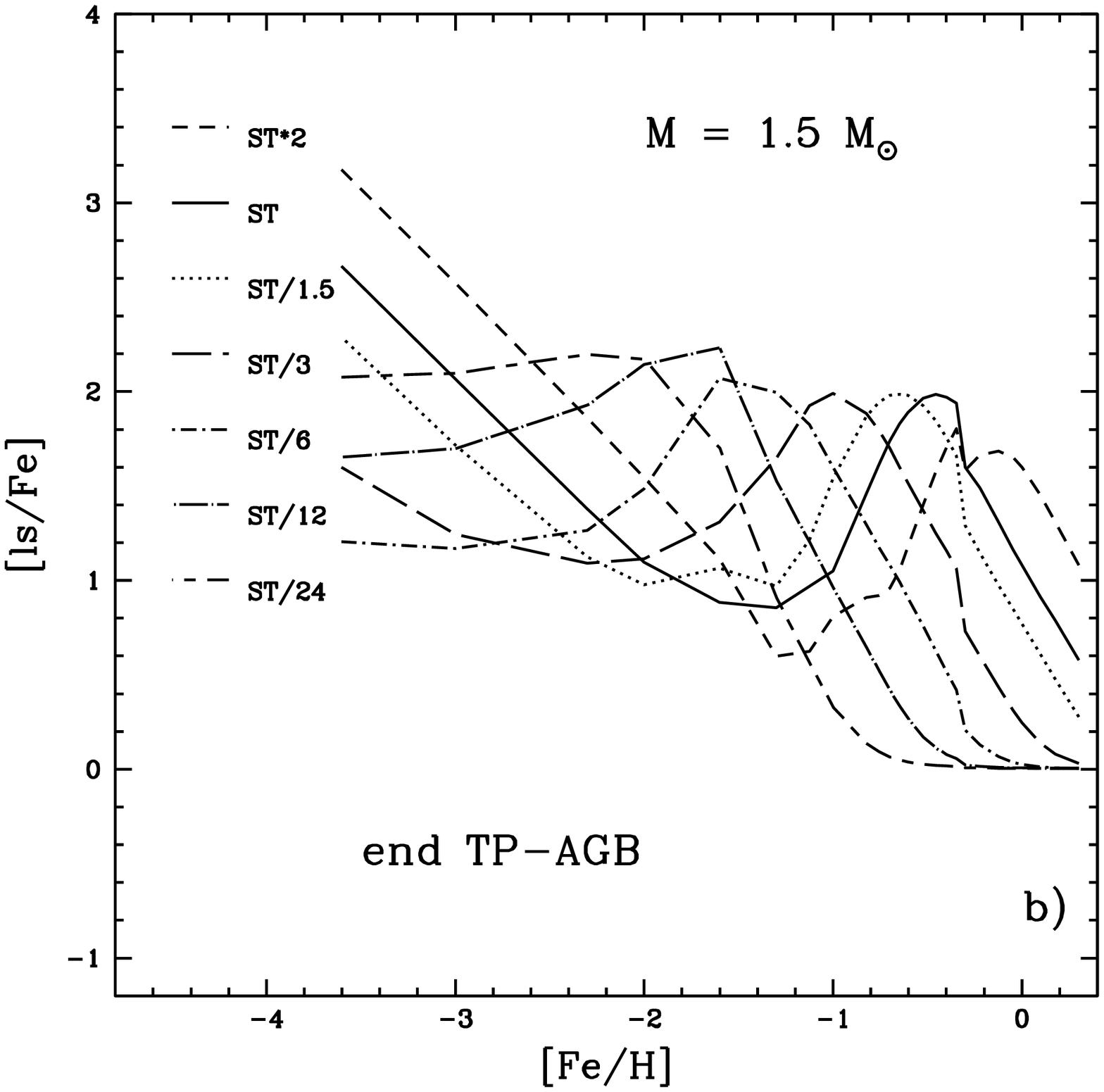}

\newpage

\plotone{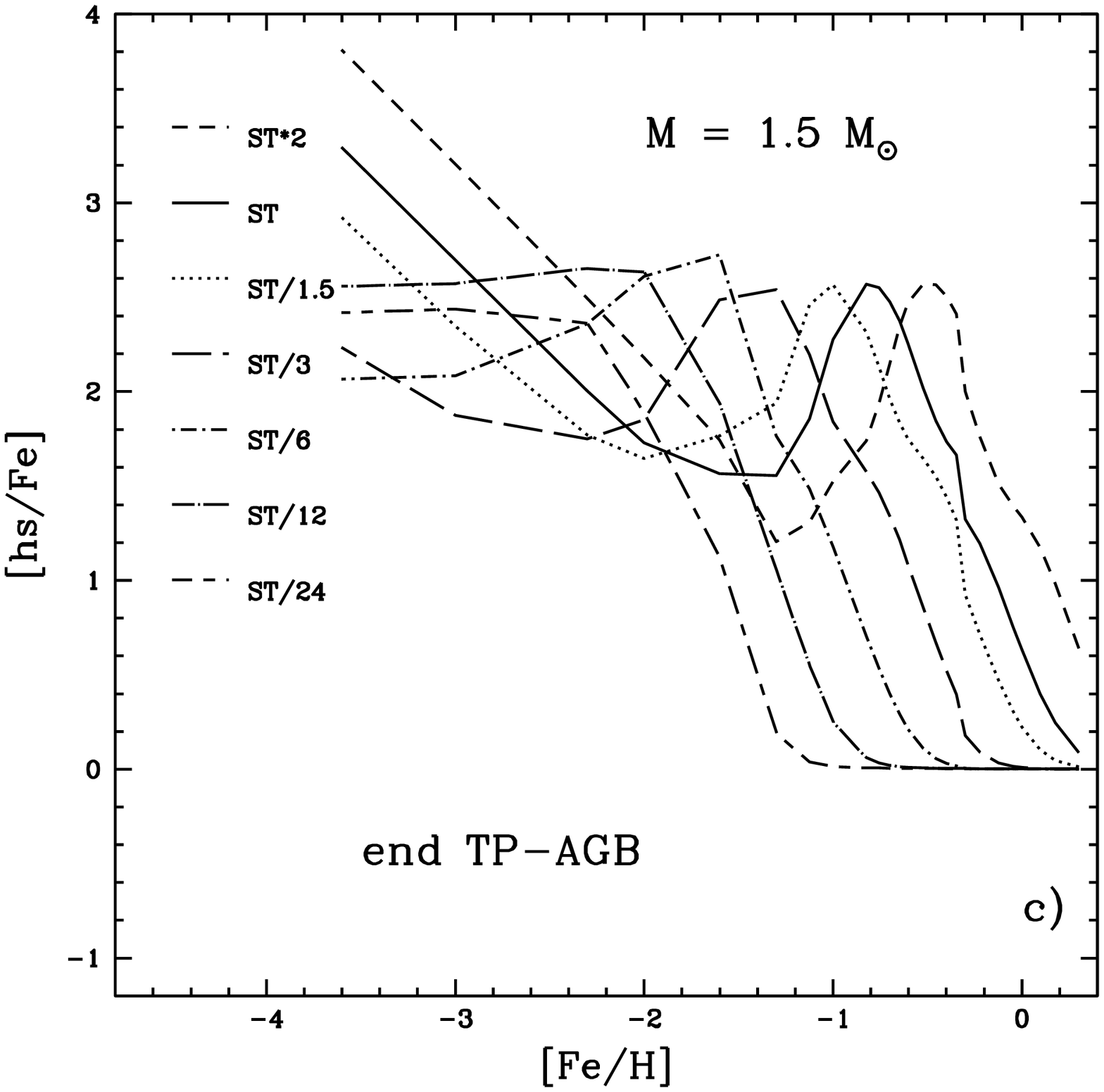}
\figcaption{{\bf (a, b, c)}. Predicted ratios [hs/ls] (a),
[ls/Fe] (b), and
[hs/Fe] (c), from AGB stellar models of 1.5 \ms. Each curve
refers to a different choice of the \ctb pocket, as shown by the
labels and discussed in the text.}

\newpage

\plotone{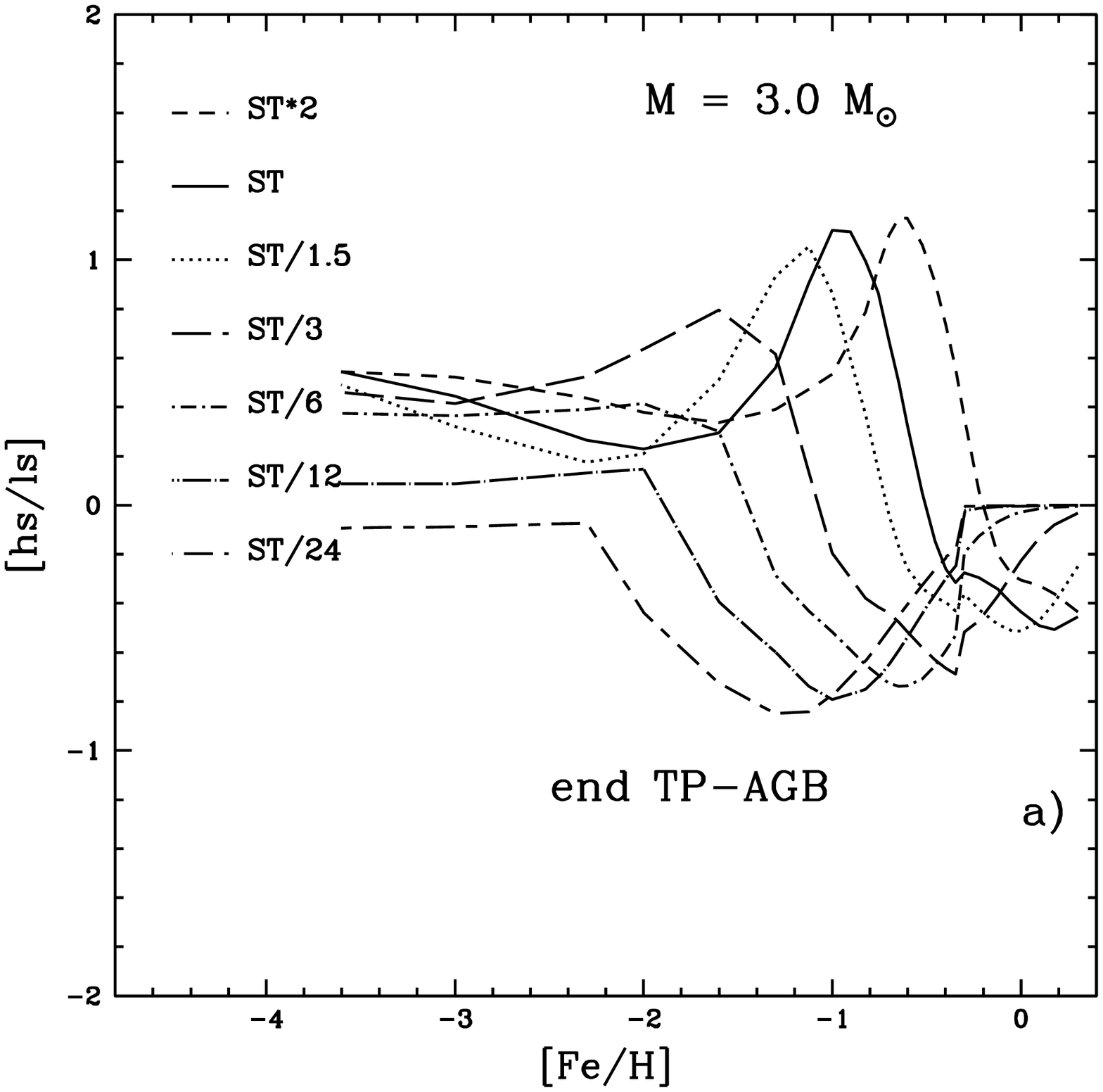}

\newpage

\plotone{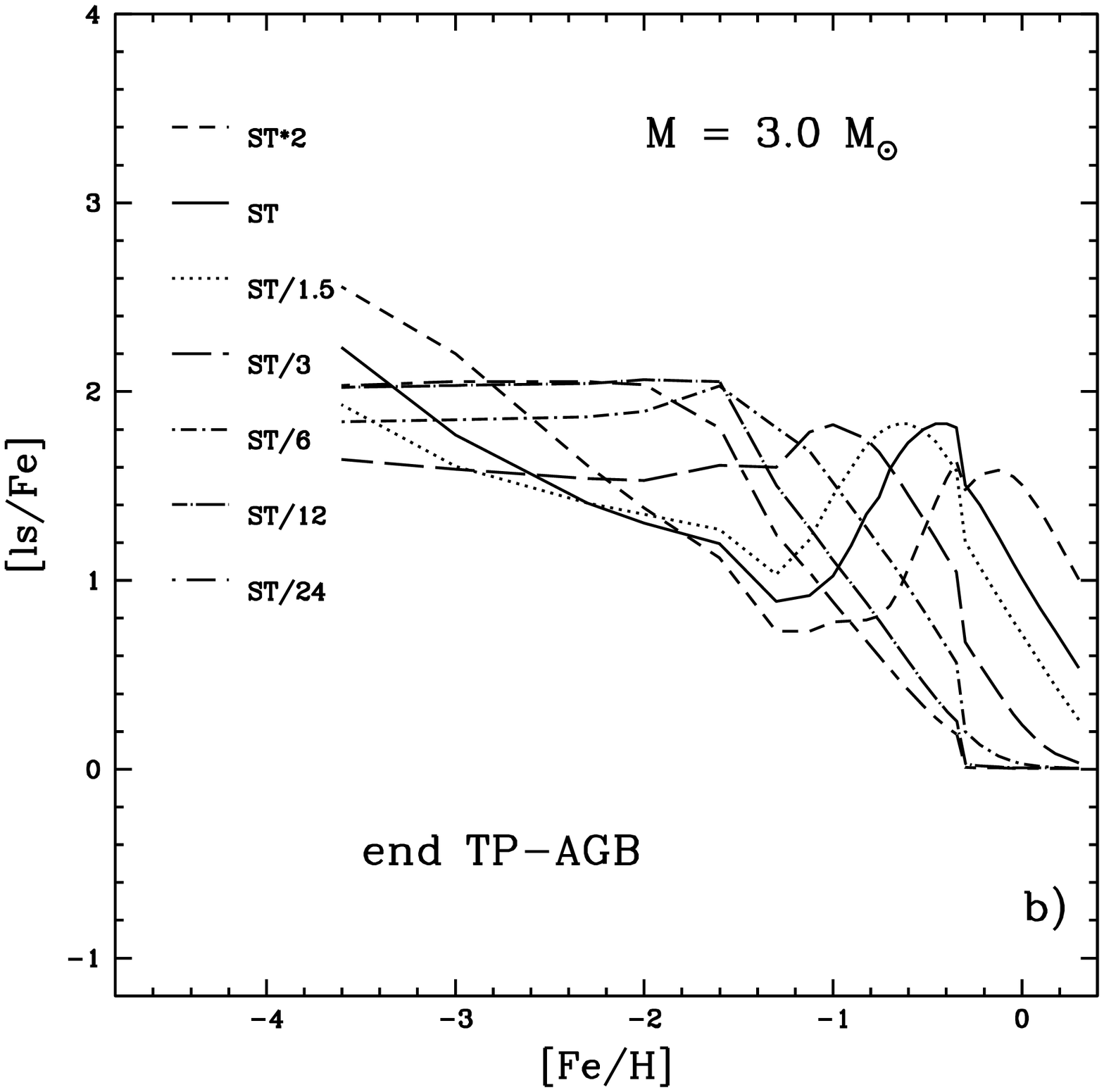}

\newpage

\plotone{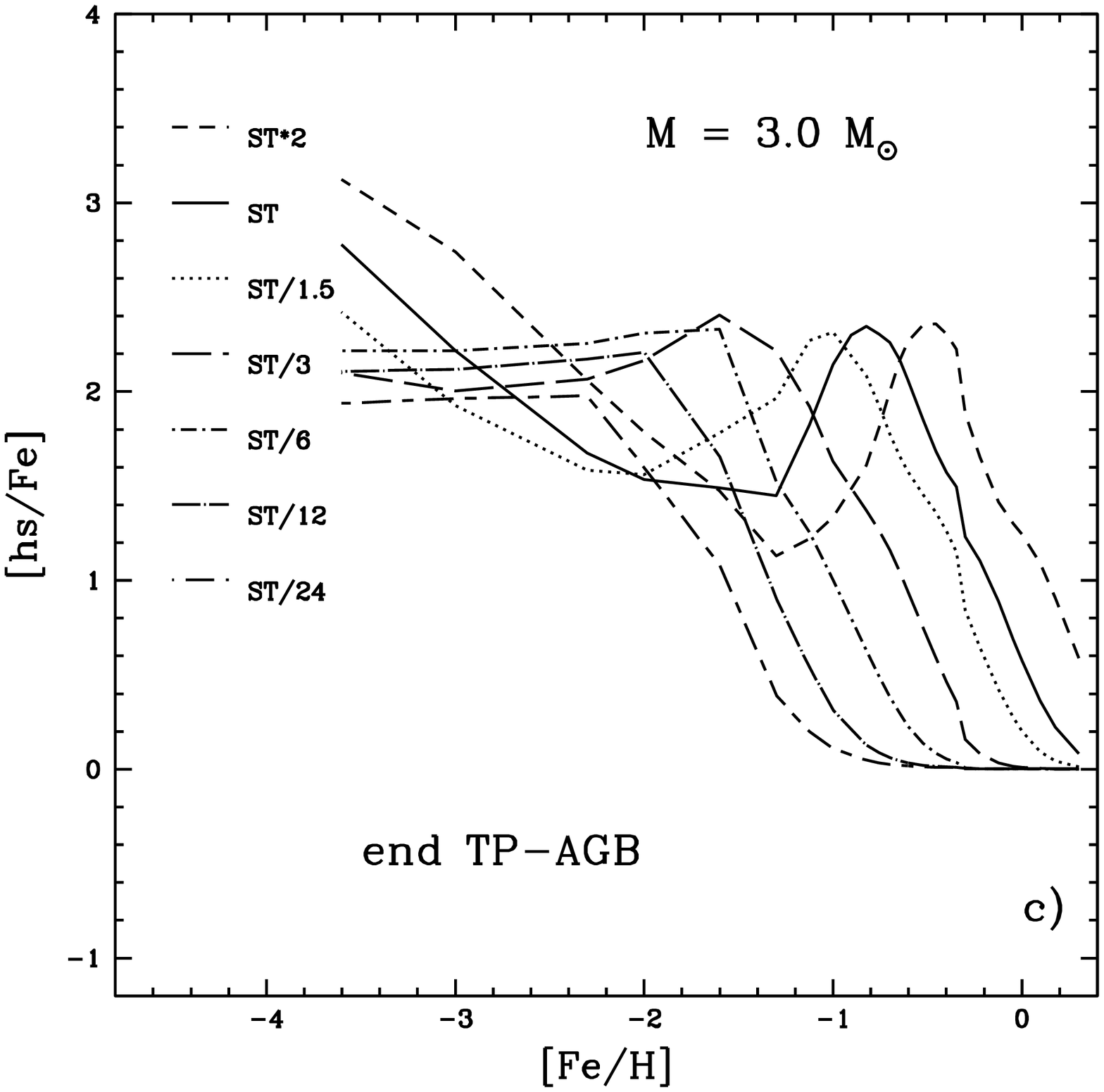}
\figcaption{{\bf (a, b, c)}. Same as Figure 3, but for 3 \msb stellar
models.}

\begin{figure}

\newpage

\centerline{\psfig{file=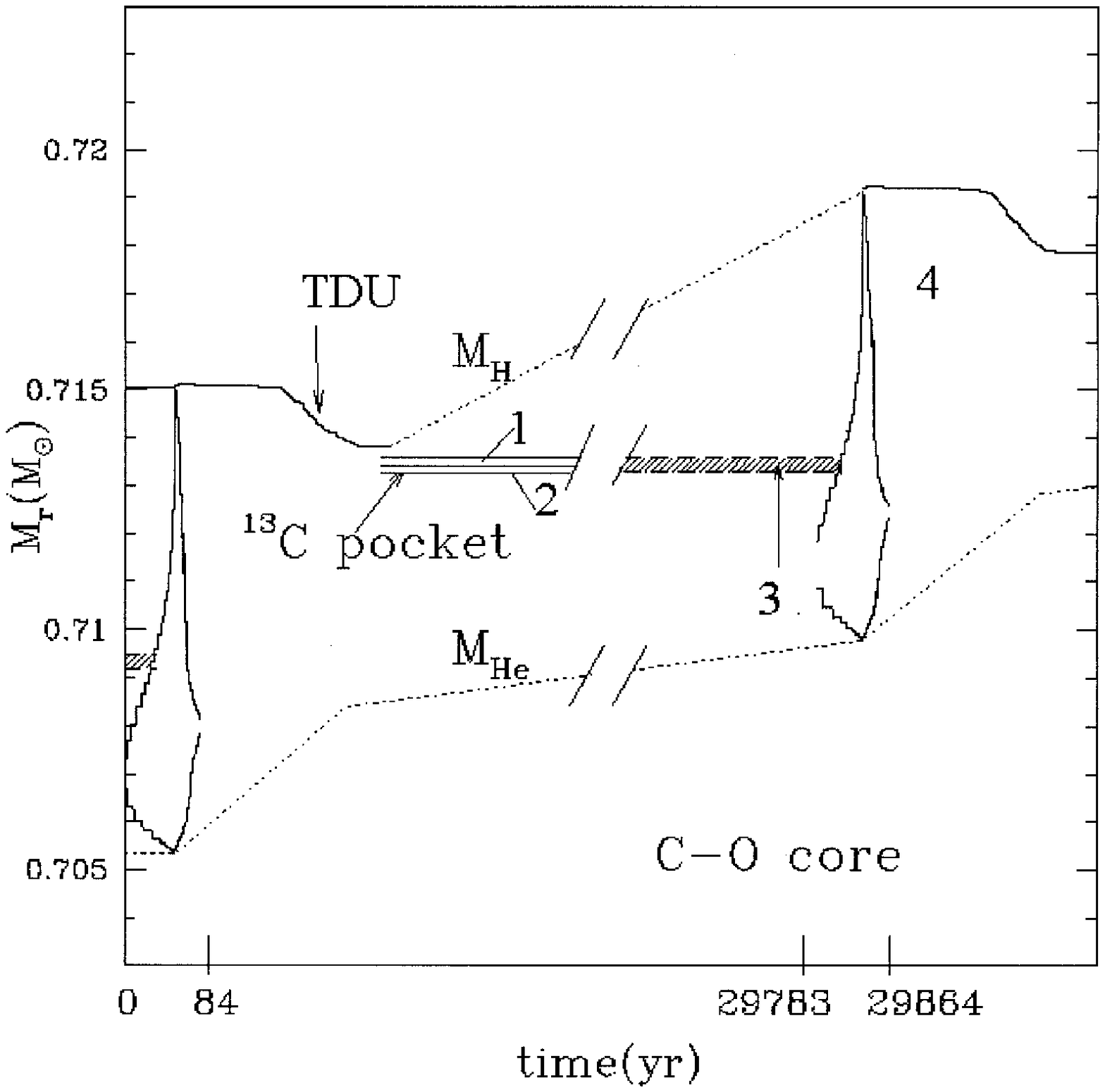,angle=-180}}
\caption[h]{A sketch of the He-intershell structure through the
development of two subsequent thermal pulses, showing the region
where the \ctb pocket forms. The zones relevant for understanding
the results listed in Table 1 are indicated.}
\end{figure}

\newpage

\plotone{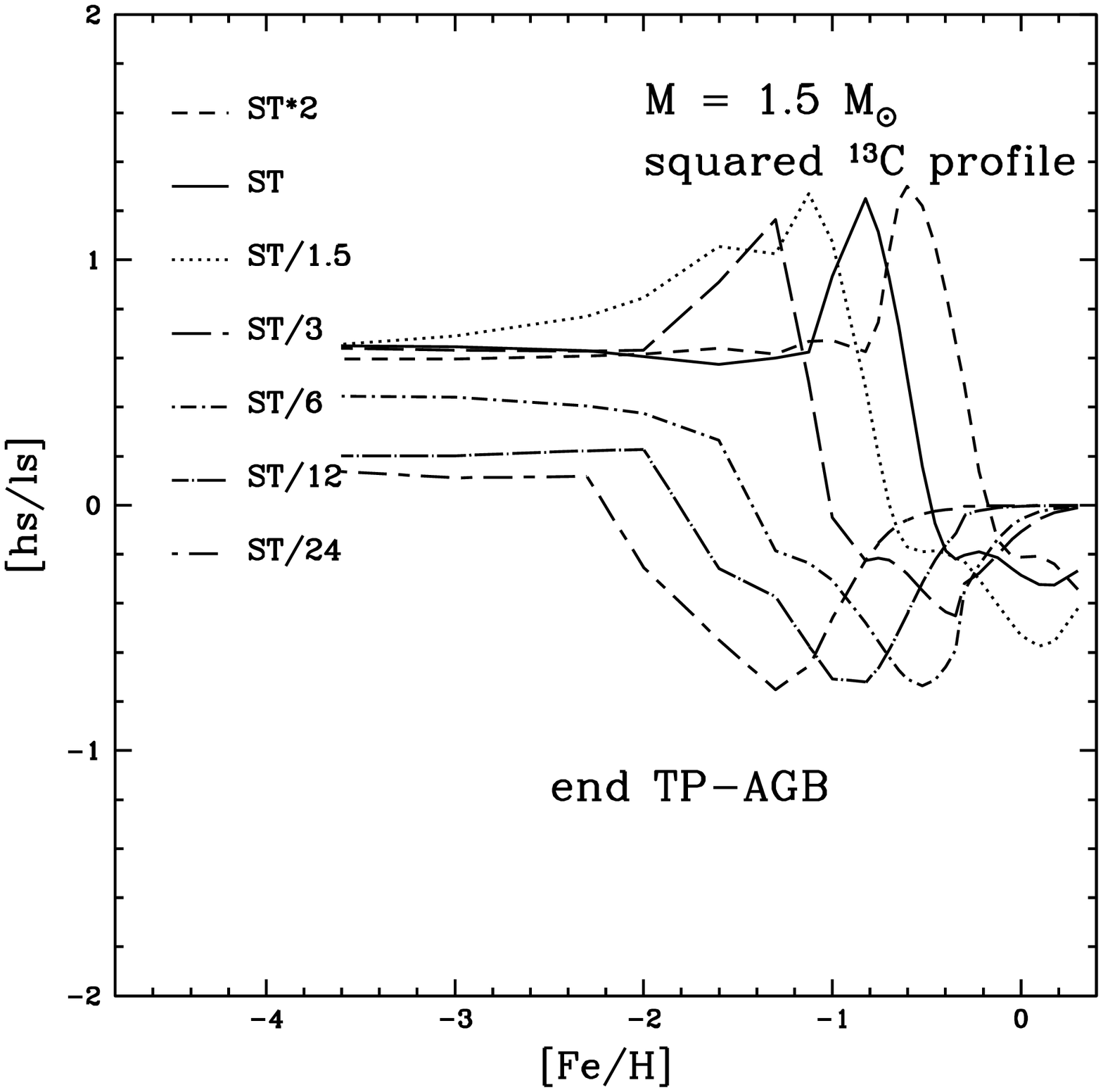}
\figcaption{Predicted [hs/ls] ratios by AGB stellar models of 1.5
\msb versus [Fe/H] for a flat \ctb profile in the pocket, as
discussed in the text.}

\newpage

\plotone{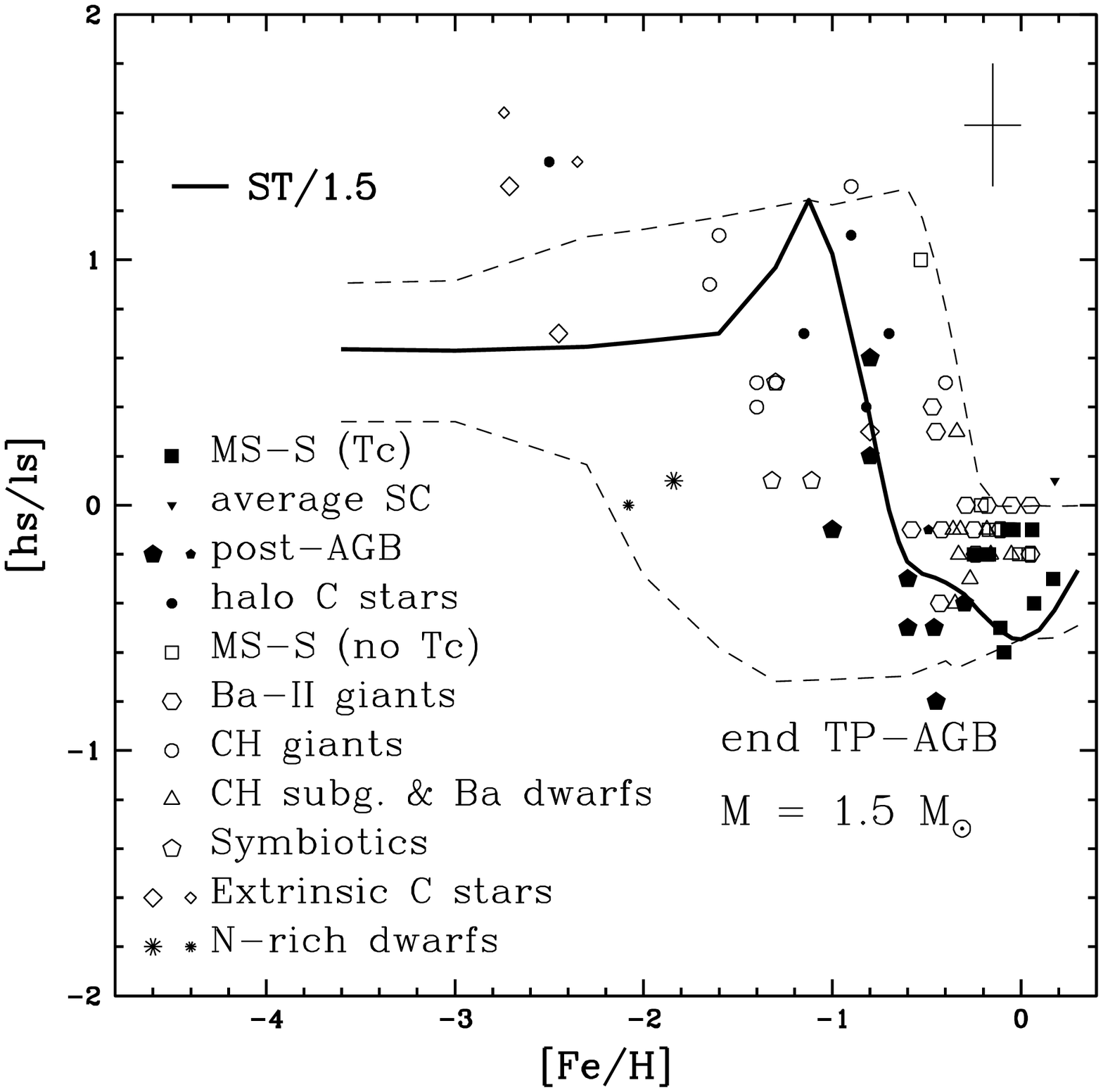}
\figcaption{Observed abundance ratios [hs/ls] in the chosen set of
intrinsic and extrinsic AGB stars at all metallicities. Filled
and unfilled symbols refer to intrinsic and extrinsic stars,
respectively. The typical value of the observational uncertainty
is shown. SC stars have been represented
by their average (see Table 2a). Small symbols refer to stars
that have lower weight in our analysis, as explained in the text.
Curves represent the predictions from the models specified on the
figure itself.}

\newpage

\plotone{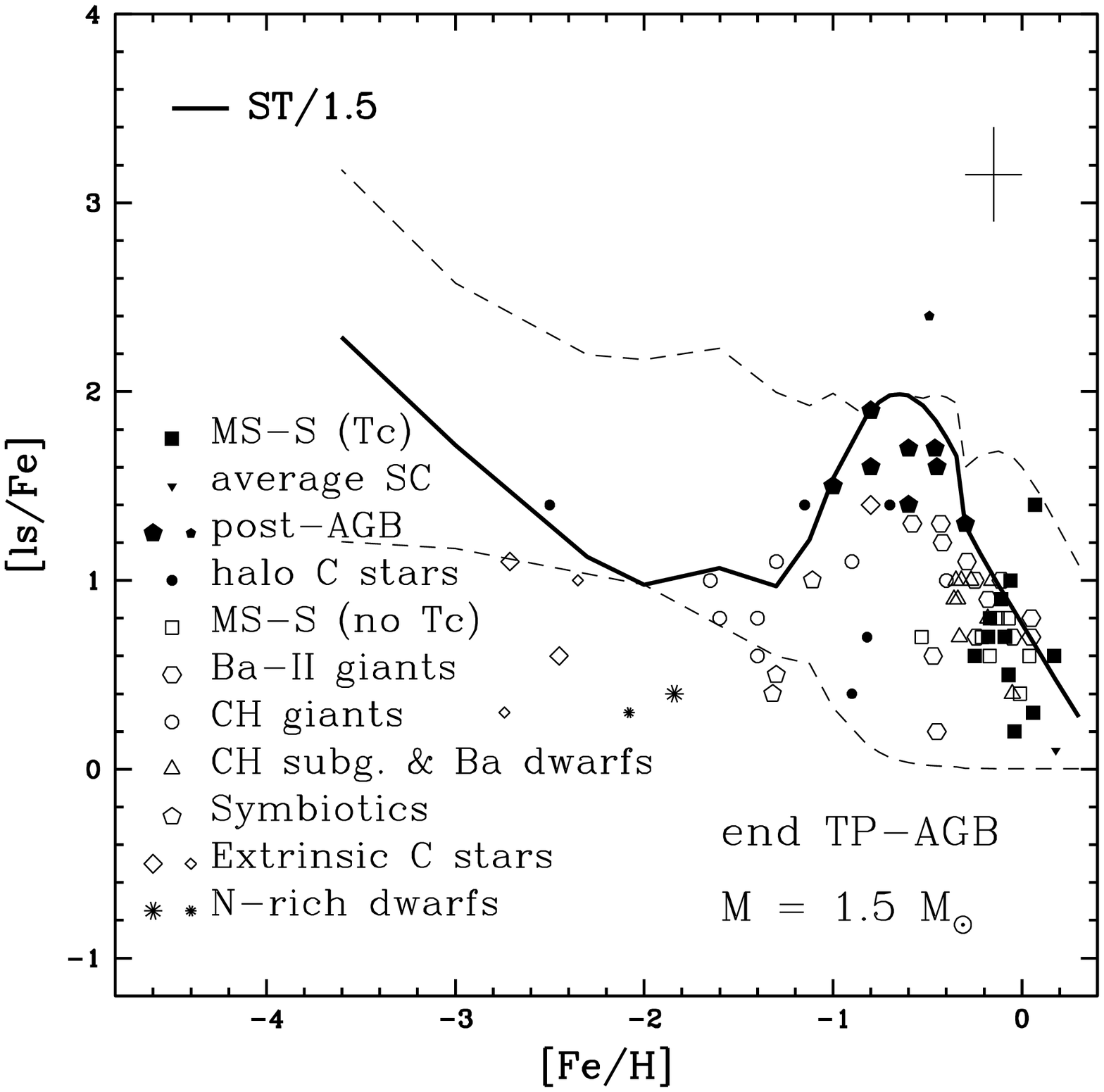}
\figcaption{Same as Figure 7, for the [ls/Fe] abundance ratios.}

\newpage

\plotone{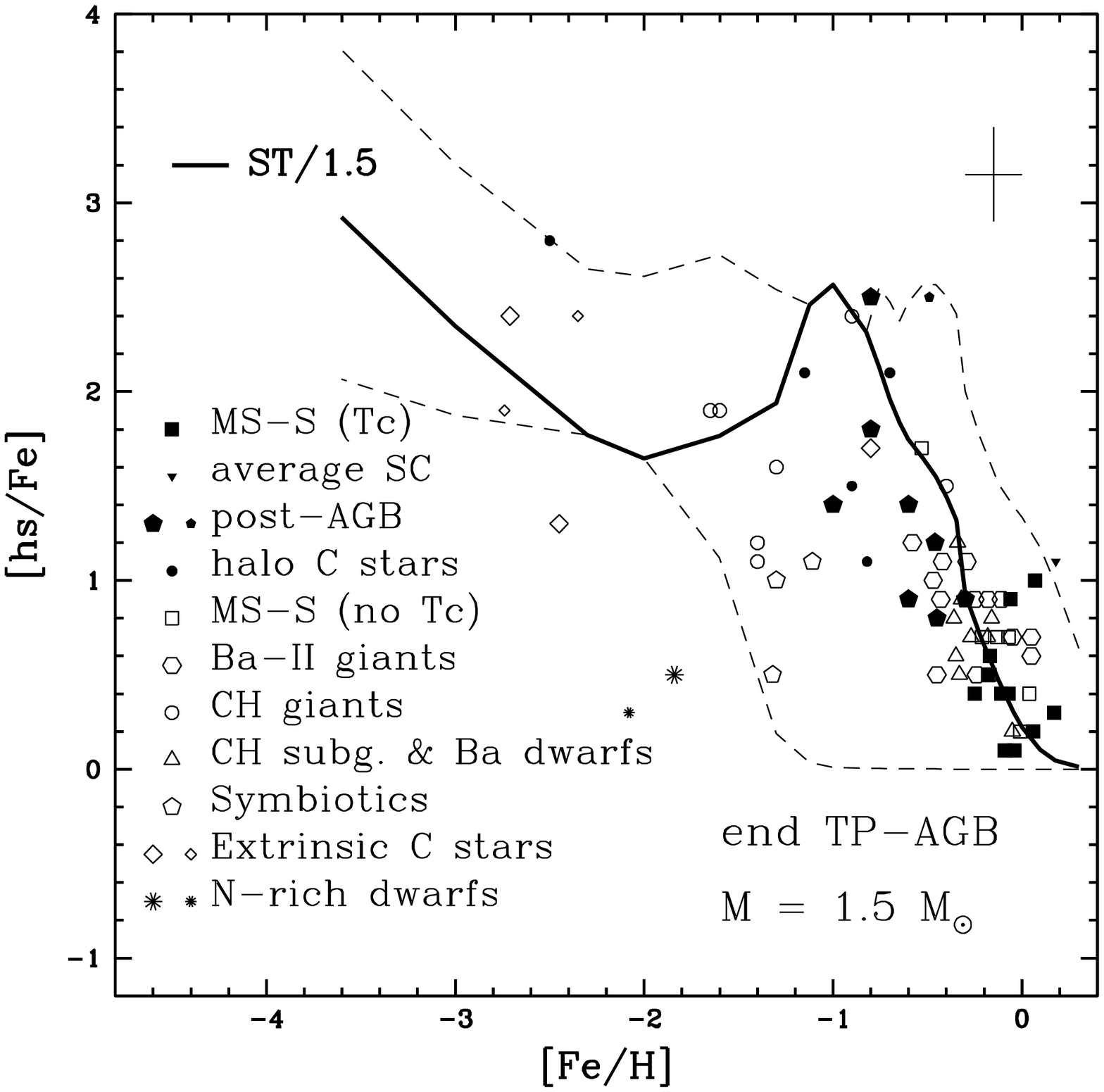}
\figcaption{Same as Figure 7, for the [hs/Fe] abundance ratios.}

\begin{figure}

\newpage

\centerline{\psfig{file=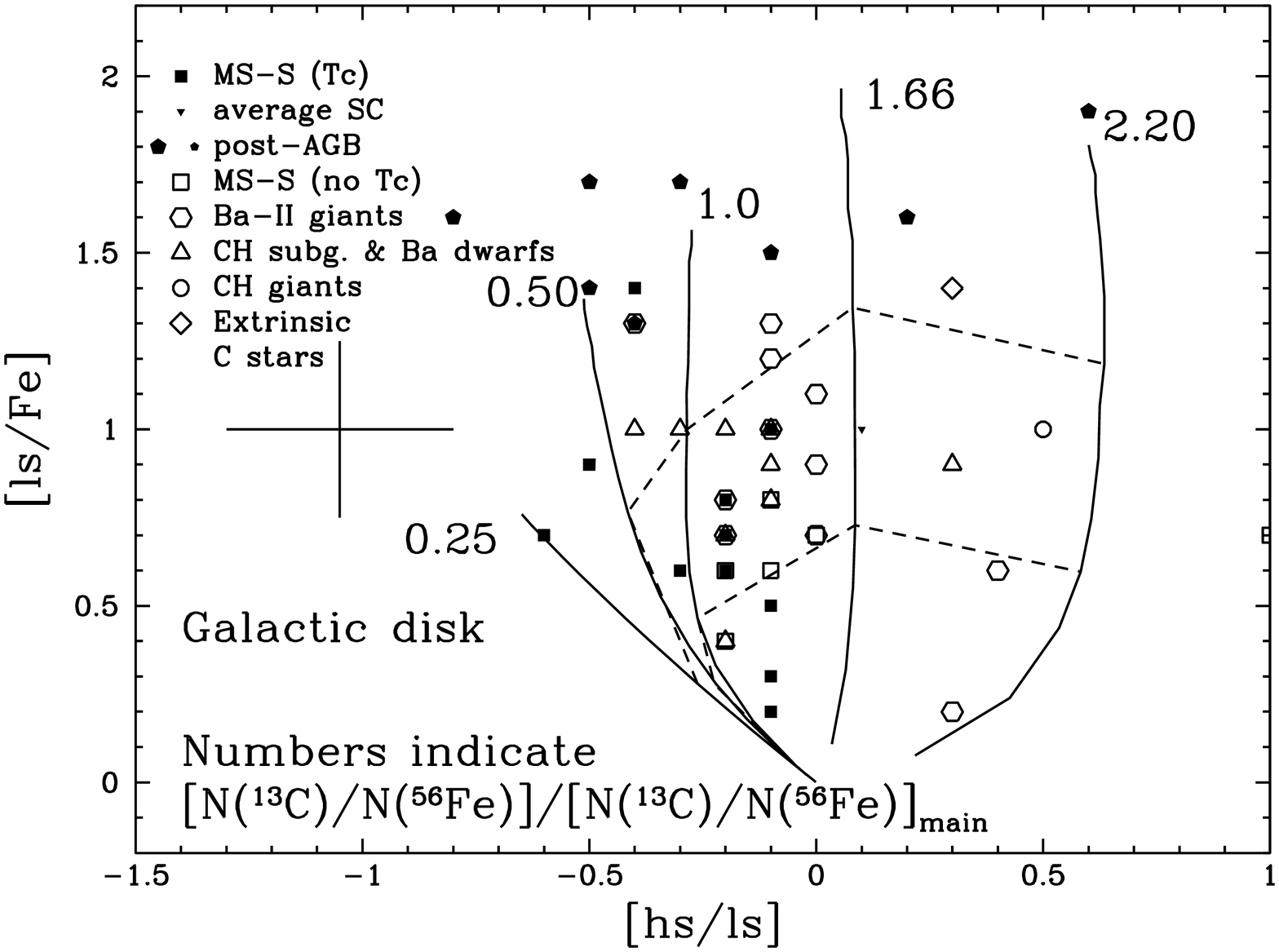,height=17cm,width=15.5cm,angle=-90}}
\caption[h]{The observed trend of the light $s$-element abundances
[ls/Fe] versus [hs/ls], for Galactic disk intrinsic and extrinsic
AGB stars. Continuous curves refer to envelope models with
different $s$-process efficiency, as monitored by the
$N$($^{13}$C)/$N$($^{56}$Fe) ratio (here normalized to the case
that fits the main component in the solar system). Dashed lines
connect the points corresponding to the 4th and 8th dredge up
episode, to make clear how the stars distribute along the TP-AGB
evolutionary sequence.}
\end{figure}

\newpage

\plotone{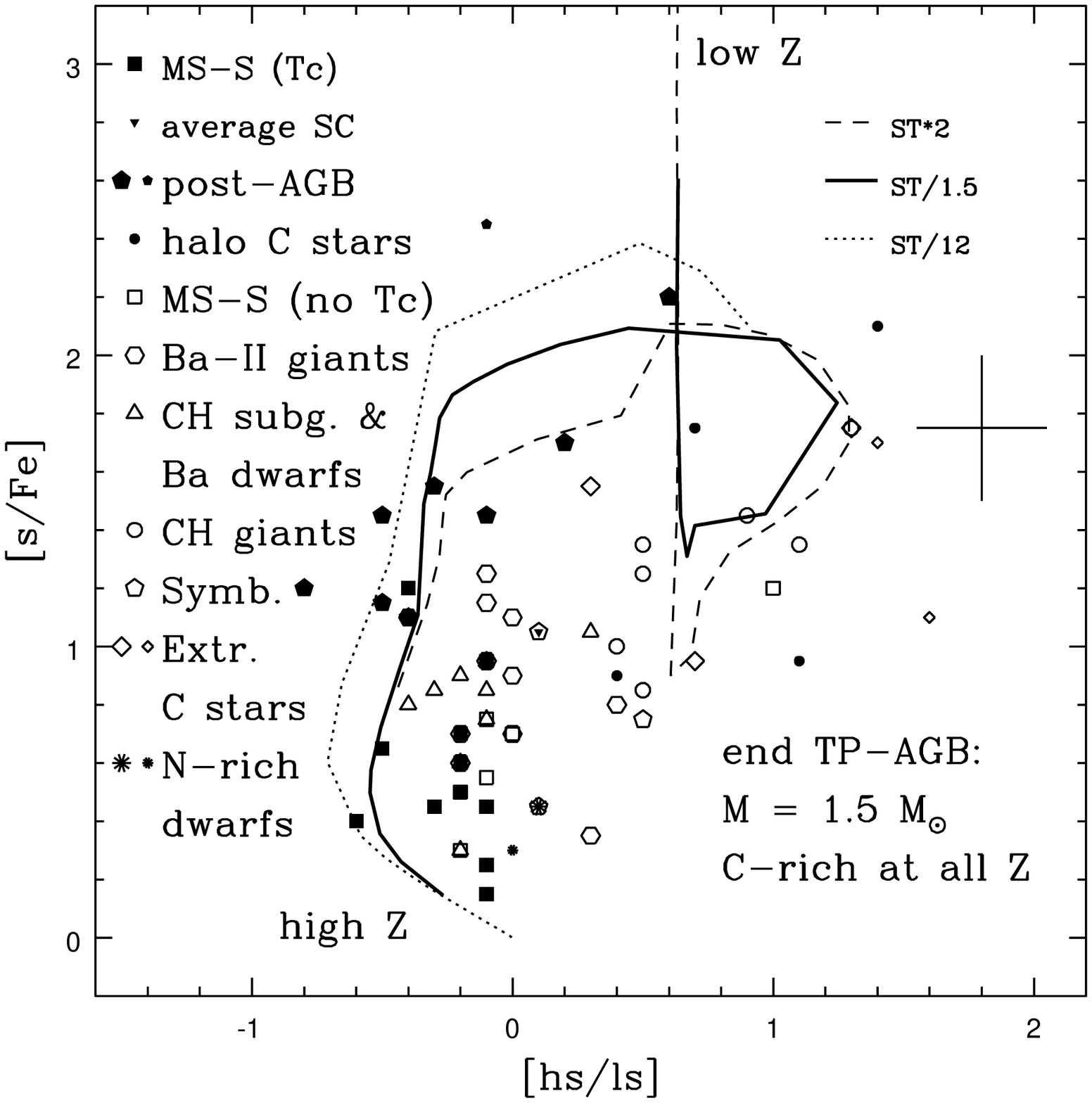}
\figcaption{The average $s$-process enrichment [s/Fe] as a
function of the [hs/ls] abundance ratio, for the various classes
of stars studied. The post-AGB supergiants lay in the region of
the maximum $s$ enrichment, as expected. Curves refer to 1.5 \msb
AGB model stars. They reflect the enrichment of $s$ elements for
different metallicities at the very end of the TP-AGB phase. The
regions where the highest and lowest metallicities explored lie
are indicated.}

\newpage

\plotone{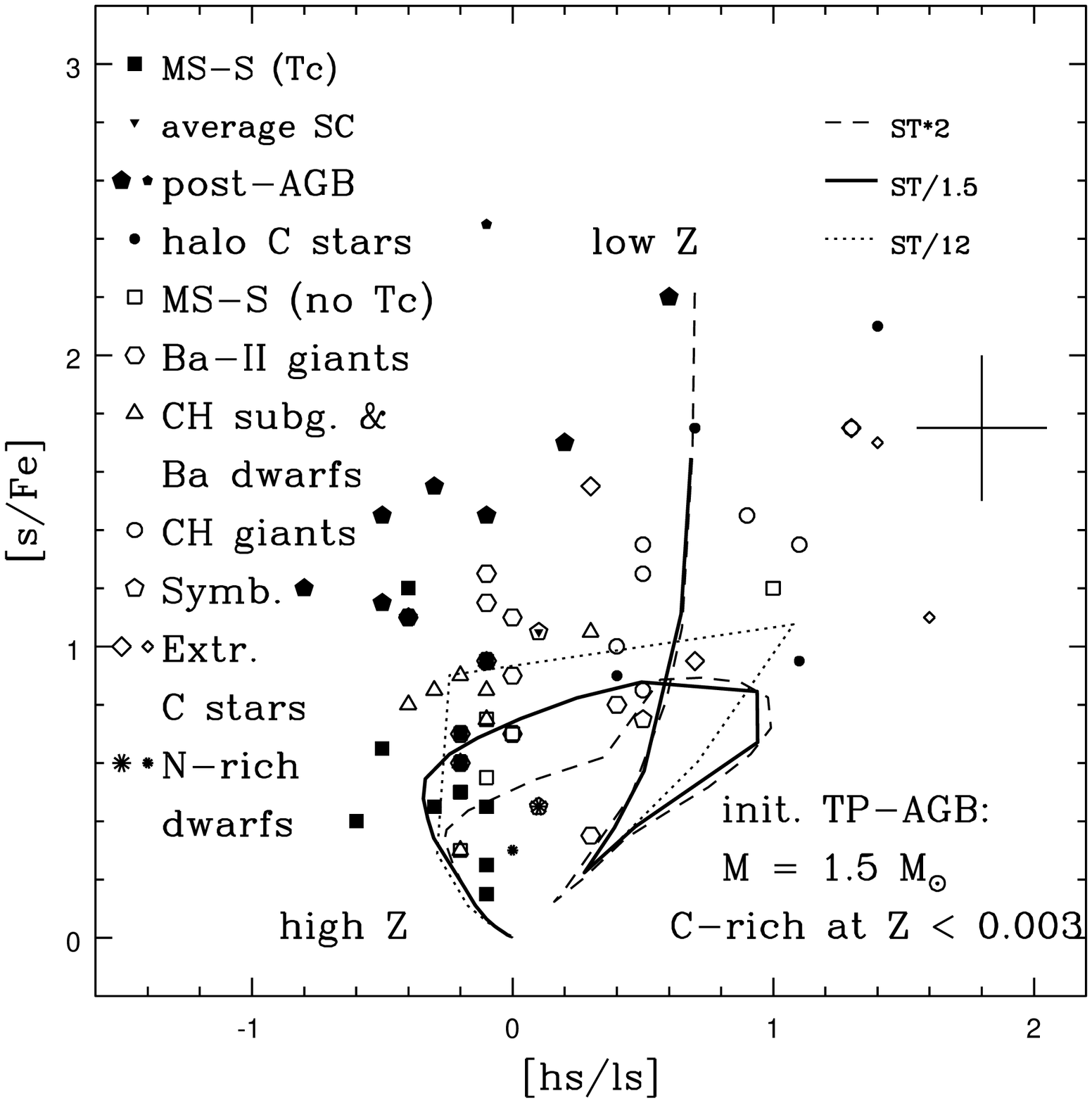}
\figcaption{Same as Figure 11, but for a much larger dilution
with an
unprocessed envelope, as obtained at an initial phase of TP-AGB
evolution, after only 4 TDU episodes.}

\newpage

\plotone{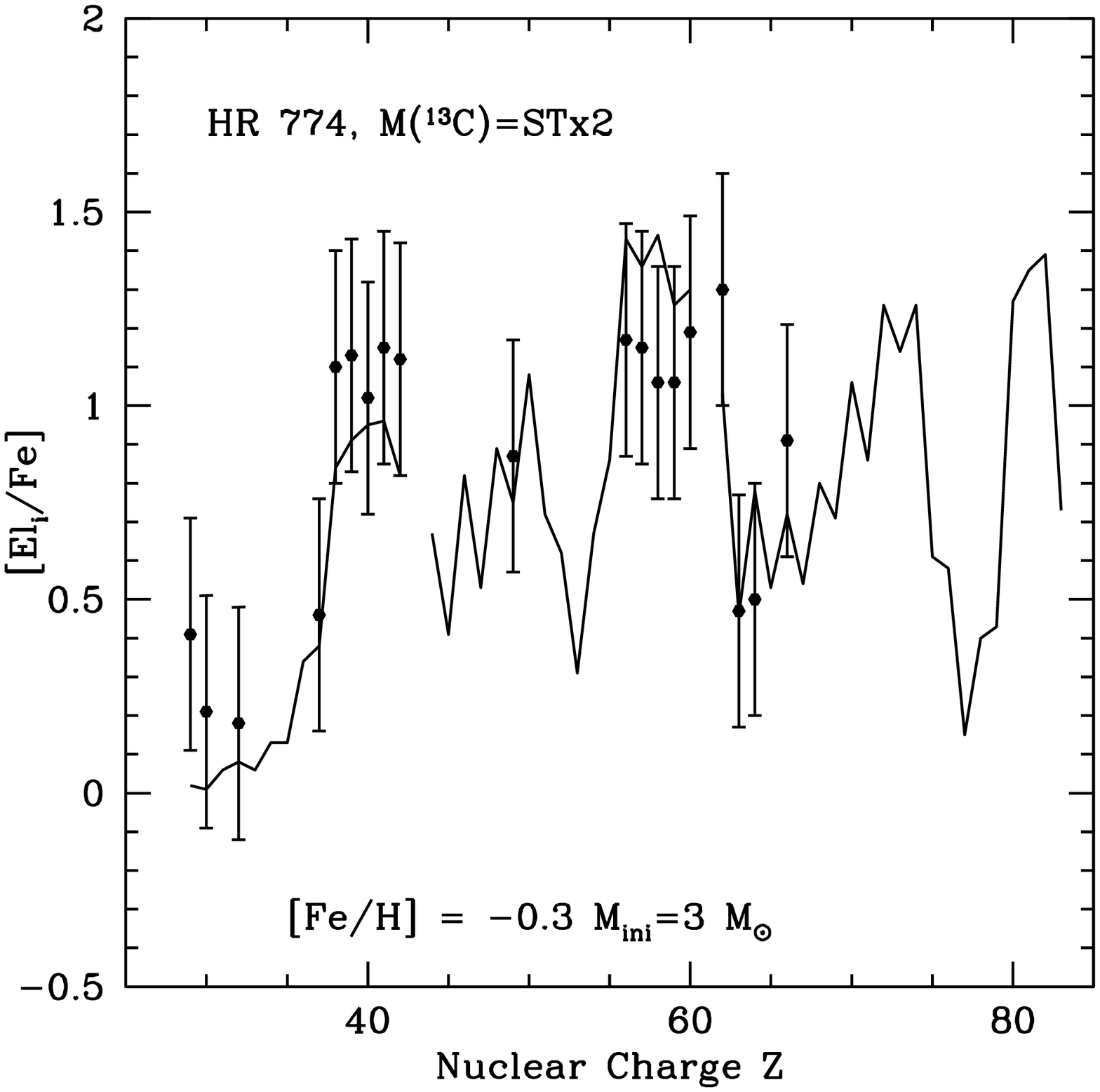}
\figcaption{A detailed reproduction of observed abundances for
the classical Ba II giant HR~774. Observations are from Smith
(1984) and Tomkin \& Lambert (1983), as compiled in Paper II. See
text for details.}

\newpage

\plotone{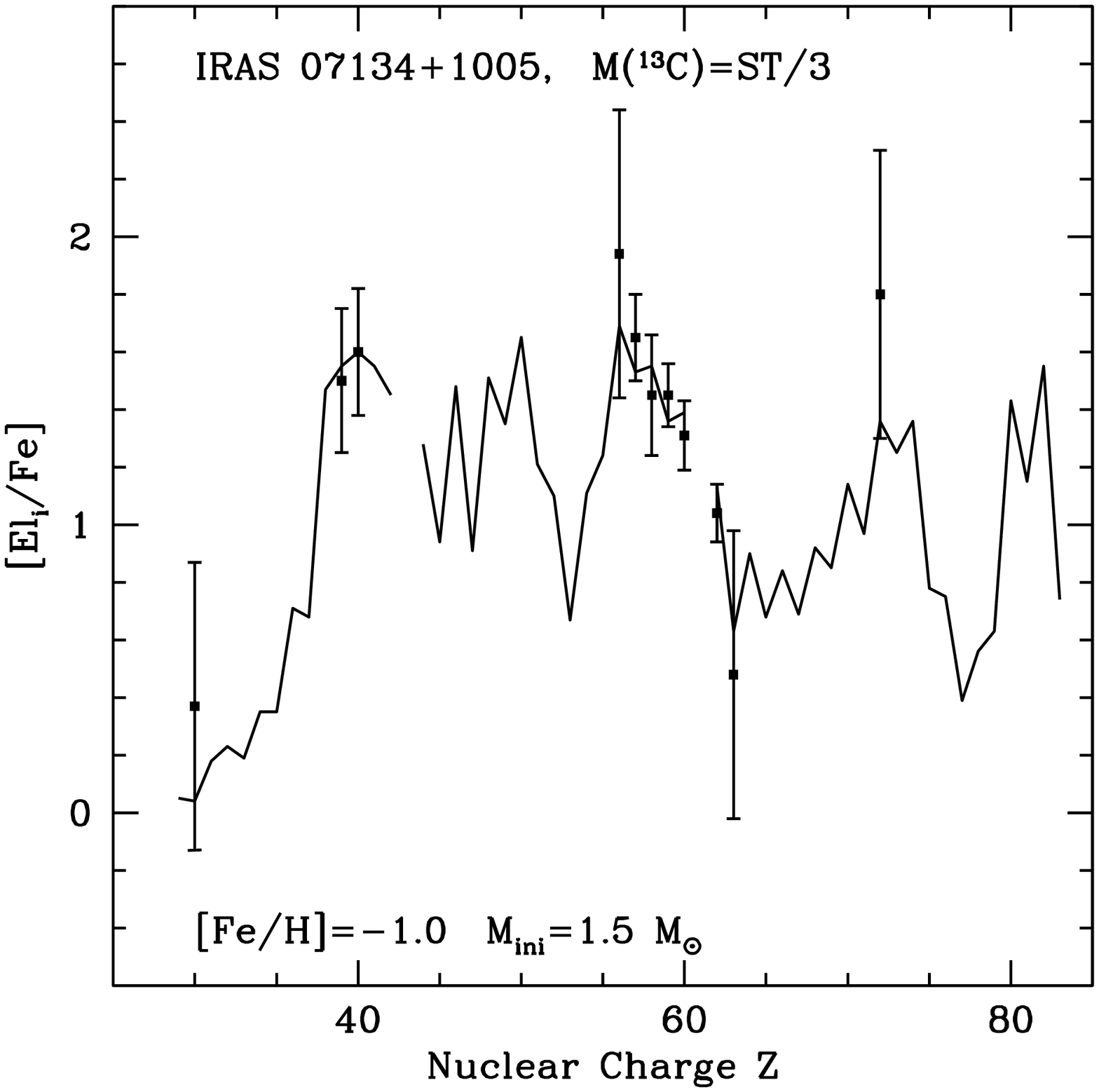}
\figcaption{A detailed reproduction of observed abundances
for the post-AGB supergiant IRAS~07134+1005. Observations are from Van
Winckel \& Reyniers (2000). See text for details.}

\newpage

\plotone{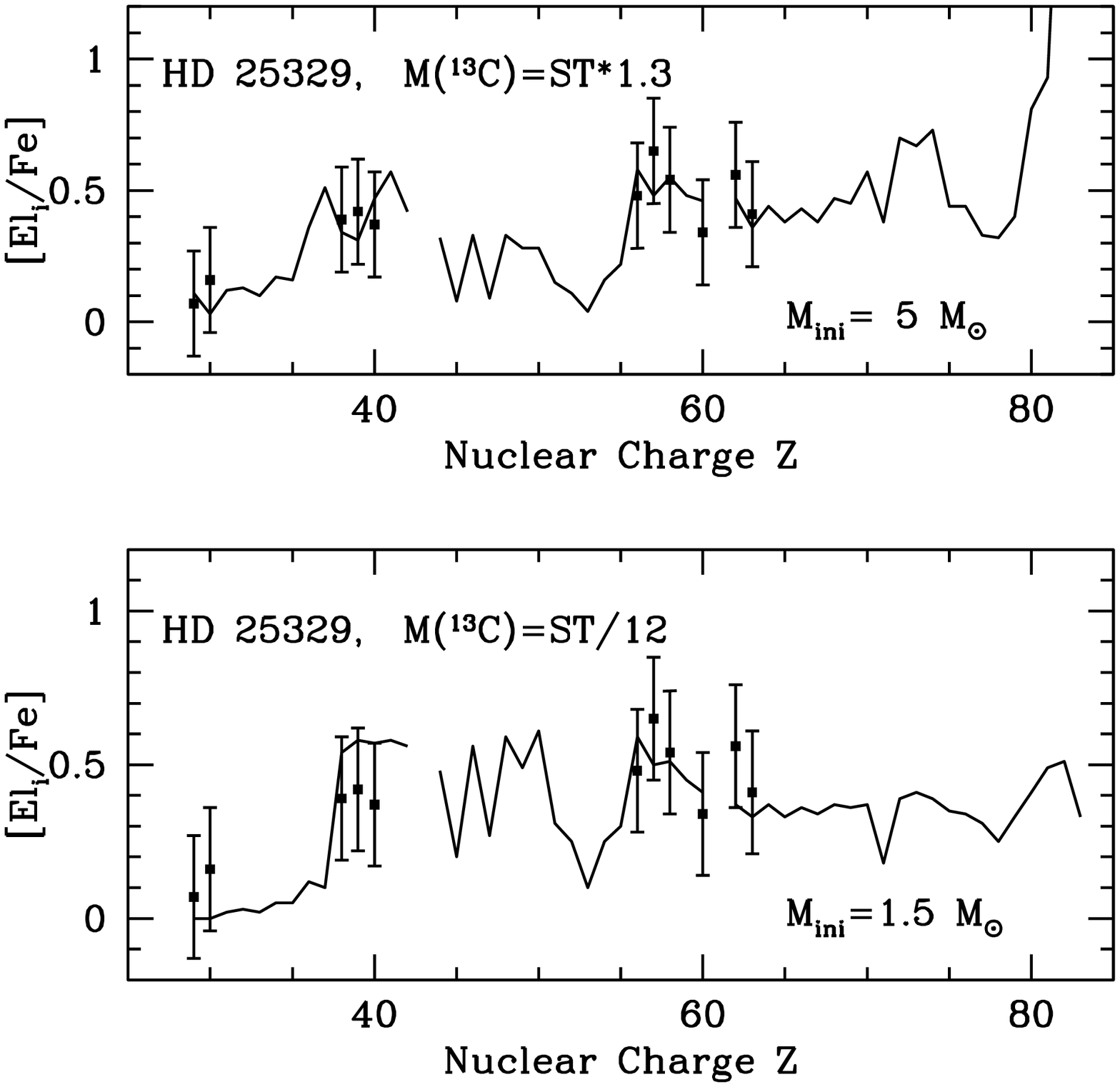}
\figcaption{Two attempts at reproducing the observed data for
the population II $s$-enriched dwarf HD~25329, performed with a
LMS and a IMS model. Observations are from Beveridge \& Sneden
(1994). See text for details.}

\newpage

\begin{table}
\begin{center}
{\bf {\large T}ABLE 1}\\
\vspace{0.5cm}
\begin{tabular}{|r|r|r|r|r|r|r|}
\hline
 & & & & & & \\
TP &  zone   & M(TP) &   $\delta$($\tau$)  &
$^{94}$Zr/$^{94}$Zr$_{\odot}$ & $^{138}$Ba/$^{138}$Ba$_{\odot}$ &
$^{208}$Pb/$^{208}$Pb$_{\odot}$   \\
 &  & (M$_{\odot})$  &  (mbarn$^{-1}$)  &  &  &  \\
\hline
 1 & layer 1   &    &      0.194   & 266   &          7.5  &          3.8 \\
   & layer 2   &    &      0.233   & 1240   &         15.9  &          4.7
\\
\hline
  & $^{13}$C pocket  &   &         &   973       &   13.6  &          4.4 \\
  &  end 2$^{\rm nd}$ TP    & 0.019    &     &   39   &          1.5  &
1.1 \\
\hline \hline
 2 & layer 1   &   &       0.189   &      1020   &        329    &
4.1 \\
 &   layer 2   &   &       0.229   &      1920   &        684    &
5.4 \\
\hline
 & $^{13}$C pocket &  &            &      1670   &        587    &
5.0 \\
 &   end 3$^{\rm rd}$ TP    & 0.018    &   &    91   &           25    &
1.2 \\
\hline \hline
 5 & layer 1   &   &       0.183   &      1680   &         730    &
19  \\
   & layer 2   &   &       0.222   &      2540   &       1330    &
34  \\
\hline
  & $^{13}$C pocket &  &           &      2300   &       1170    &
30 \\
 &  end 6$^{\rm th}$ TP      &  0.016 &     &  175    &          82     &
2.7 \\
\hline \hline 10 & layer 1  &   &        0.177   &      1880
&        986    &
68 \\
 &   layer 2  &   &        0.216   &      2720   &       1630    &
107 \\
\hline
 & $^{13}$C pocket  &   &          &      2490   &       1450    &
96  \\
 &   end 11$^{\rm th}$ TP  & 0.013     &           &       236    &  145     &
10  \\
\hline \hline 15 &  layer 1 &    &       0.168   &      1960
&       1030    &
81 \\
 &    layer 2 &    &       0.207   &      2770   &       1670    &
123 \\
\hline
 &  $^{13}$C pocket &    &         &      2550   &       1490    &
112 \\
 &    end 16$^{\rm th}$ TP  & 0.011    &  &      273    &        174     &
14  \\
\hline \hline 20 & layer 1  &    &       0.161   &      2160
&       1170    &
99 \\
 &   layer 2  &    &       0.199   &      2960   &       1830    &
149 \\
\hline
 & $^{13}$C pocket &    &          &      2740   &       1650    &
136 \\
 &    end 21$^{\rm th}$ TP    & 0.0093  &          &      351    &       231  &
20  \\
\hline \hline
\end{tabular}
\end{center}
\end{table}

\newpage
\begin{table}
\begin{center}
{{\bf {\large T}ABLE 2a.} {\large I}NTRINSIC {\large
G}ALACTIC-DISK {\large AGB} {\large S}TARS} \\
\vspace{0.5cm}
\begin{tabular}{|l|l|r|r|r|r|c|l|}
\hline
STAR & ALIAS & [Fe/H] & [ls/Fe] & [hs/Fe] & [hs/ls] & Type & Ref. \\
\hline HD~30959 & o$^1$~Ori  &  $-$0.11  &    0.9   & 0.4    &
$-$0.5   &   MS/S (Tc) & (1) \\
HD~58521 & Y~Lyn    &  $-$0.18  &    0.7     &   0.5      &
$-$0.2     & idem & (2) \\
HD~64332 & NQ Pup    &  $-$0.25  &    0.6   & 0.4    &
$-$0.2 &    idem & (2) \\
HD~78712 & HR~3639      &  $-$0.17  &    0.8   & 0.6    &
$-$0.2 &    idem & (3) \\
HD~106198& HR~4647     &  $-$0.04  &    0.2     &   0.1  &  $-$0.1
&    idem & (2) \\
HD~163990 & HR~6702      &  $-$0.09  &    0.7   & 0.1    & $-$0.6
& idem & (1) \\
HD~172804   & V679~Oph &  $-$0.06  &    1.0   & 0.9    &    $-$0.1
&    idem & (2) \\
HD~199799    & V2141~Cyg  &    0.17  &    0.6   & 0.3    & $-$0.3
& idem & (2) \\
HD~200527 & HR~8062      &   0.06  &    0.3   & 0.2    & $-$0.1
& idem & (3) \\
HD~216672 & HR~8714      &  $-$0.07  &    0.5   & 0.4    & $-$0.1
& idem & (3) \\
BD~+48$^\circ$1187&TV Aur & 0.07  &  1.4 &   1.0      &  $-$0.4
& idem & (2) \\
\hline HD~144578 & RR~Her       &   0.20  &    1.0   &     1.5
&     0.5   & SC & (4) \\
HD~198164 & CY~Cyg       &   0.20  &    0.8   &     0.7    &
$-$0.1   & idem & (4) \\
HD~286340 & GP~Ori       &   0.00  &    1.4   & 1.2    & $-$0.2
& idem & (4) \\
CD~$-$52$^\circ$5798 & AM~Cen       &   0.30  &    0.9   & 1.0
&  0.1 & idem & (4) \\
\hline Average SC & (see text) &   0.18  &    1.0   &     1.1
&    0.1    &  ---   & --- \\
\hline HD~56126 & IRAS~07134+1005 & $-$1.00  &    1.5   & 1.4 &
$-$0.1   & post AGB &    (5) \\
HD~187885 & IRAS~19500$-$1709 &   $-$0.60  &    1.4   & 0.9    &
$-$0.5   & idem &    (5) \\
HD~235858 & IRAS~22272+5435  & $-$0.49  &    2.4   & 2.5    &
$-$0.1 & idem &   (6) \\
IRAS~04296+3429 & GLMP~74&   $-$0.60  &    1.7   & 1.4    & $-$0.3
& idem & (5) \\
IRAS~05341+0852 & GLMP~106 & $-$0.80  &    1.9   & 2.5    & 0.6
&idem & (5) \\
IRAS~07430+1115 & GLMP~192 &  $-$0.46  &    1.7   & 1.2    &
$-$0.5& idem &   (7) \\
IRAS~22223+4327 & GLMP~1058 &   $-$0.30  &    1.3   &     0.9 &
$-$0.4 & idem &    (5) \\
IRAS~23304+6147 & GLMP~1078 &  $-$0.80  &    1.6   & 1.8    & 0.2
&idem &    (5) \\
IRAS~Z02229+6208 &--- & $-$0.45  &  1.6   & 0.8    &    $-$0.8   &
idem &   (7) \\
\hline
\end{tabular}
\end{center}
\end{table}

\newpage

\begin{table}
\begin{center}
{{\bf {\large T}ABLE 2b.} {\large I}NTRINSIC {\large
G}ALACTIC-HALO {\large AGB} {\large S}TARS} \\
\vspace{0.5cm}
\begin{tabular}{|l|l|r|r|r|r|c|l|}
\hline
STAR & ALIAS & [Fe/H] &[ls/Fe] & [hs/Fe]& [hs/ls]& Type & Ref. \\
\hline HD~25408     & UV~Cam & $-$0.82  &    0.7   & 1.1    & 0.4
& Halo C star
& (8) \\
HD~59643     & NQ~Gem & $-$0.70  &    1.4   & 2.1    &     0.7 &
idem
& (8) \\
HD~187216    & SAO~3243  & $-$2.50  &    1.4   & 2.8    &     1.4
& idem  &     (9) \\
HD~189711    & SAO~125356 & $-$1.15  &    1.4   & 2.1    &     0.7
&    idem
& (8) \\
HD~197604    & CGCS~4947 & $-$0.90  &    0.4   & 1.5    &     1.1
& idem & (8) \\
\hline
\end{tabular}
\end{center}
\end{table}

\newpage

\begin{table}
\begin{center}
{{\bf {\large T}ABLE 2c.} {\large E}XTRINSIC {\large
G}ALACTIC-DISK {\large AGB} {\large S}TARS} \\
\vspace{0.5cm}
\begin{tabular}{|l|l|r|r|r|r|c|l|}
\hline
STAR & ALIAS & [Fe/H] &[ls/Fe] & [hs/Fe]& [hs/ls]& Type & Ref. \\
\hline
HD~7531 & HR~363 & $-$0.13 & 0.8 & 0.7 & $-$0.1 & MS-S no Tc & (1) \\
HD~22649 & HR~1105 & $-$0.07 & 0.8 & 0.7 & $-$0.1 & idem & (3) \\
HD~35155 & IRAS~05199$-$0842 & $-$0.53 & 0.7 & 1.7 & 1.0 & idem & (2) \\
HD~49368 & IRAS~06457+0535 & $-$0.21 & 0.7 & 0.7 & 0.0 & idem & (2) \\
HD~119667 & BD~$-$02$^\circ$3726 & $-$0.17 & 0.6 & 0.5 & $-$0.1 & idem & (2) \\
HD~147923 & BD~+57$^\circ$1671 & $-$0.01 & 0.4 & 0.2 & $-$0.2 & idem & (2) \\
HD~151011 & BD~$-$18$^\circ$4320 & 0.04 & 0.6 & 0.4 & $-$0.2 & idem & (2) \\
\hline HD~16458 & HR~774       &   $-$0.43  &    1.3   & 0.9    &
$-$0.4   &  Ba II giant &  (10) \\
HD~44896 & SAO~196752    &   $-$0.25  &    1.0   & 0.9    &
$-$0.1   & idem &  (11) \\
HD~46407 & HR~2392      &   $-$0.42  &    1.2   & 1.1    &
$-$0.1 &    idem &  (12) \\
HD~60197     & SAO~173956 &   $-$0.05  &    0.7   & 0.7    & 0.0
& idem &  (11) \\
HD~104979    & o~Vir  & $-$0.47  &    0.6   & 1.0    &     0.4 &
idem &  (13) \\
HD~116713 & HR~5058 &   $-$0.29  &    1.1   & 1.1    & 0.0
&idem &  (12) \\
HD~121447   & IT~Vir &   0.05  &    0.7   & 0.7    &     0.0 &
idem &  (11) \\
HD~139195 & HR~5802  &   $-$0.24  &    0.7   & 0.5    &    $-$0.2
& idem & (13) \\
HD~178717& SAO~104535 &   $-$0.18  &    0.9   & 0.9    &     0.0
&  idem & (11) \\
HD~204075 &  $\zeta$~Cap &   $-$0.11  &    1.0   & 0.9    & $-$0.1
& idem &  (14,15) \\
DM~$-$64$^\circ$4333& MFU~149 & 0.05 &    0.8   & 0.6    & $-$0.2
& idem &  (11) \\
NGC~2420-D   & --- &  $-$0.45  &    0.2   & 0.5    &     0.3   &
idem
&  (16) \\
NGC~2420-X   & --- &  $-$0.58  &    1.3   & 1.2    &    $-$0.1   &
idem
&  (16) \\
\hline HD~4395   & SAO~147438   &   $-$0.33  &    0.7   & 0.5 &
$-$0.2   &    CH subgiant &    (17) \\
HD~11377 & SAO~148062    &   $-$0.05  &    0.4   & 0.2    & $-$0.2
& idem &    (17) \\
HD~88446  & SAO~98999   &   $-$0.36  &    0.9   & 0.8    &
$-$0.1 &    idem &     (17) \\
HD~89948  & SAO~178741   &   $-$0.27  &    1.0   & 0.7    & $-$0.3
& idem &  (17) \\
HD~125079    & SAO~139848 & $-$0.16  &    1.0   & 0.8    & $-$0.2
& idem &     (17) \\
HD~182274 & SAO~104781  &  $-$0.18  &    0.8   & 0.7    &   $-$0.1
&    idem &     (17) \\
HD~204613 & SAO~33445    &   $-$0.35  &    1.0   & 0.6    & $-$0.4
& idem &   (17) \\
HD~216219  & SAO~108214  & $-$0.32  &    1.0   & 0.9    & $-$0.1 & idem &(17) \\
HD~219116    & SAO~165564  & $-$0.34  &    0.9   & 1.2    & 0.3
& idem &     (17) \\
\hline HD~26  & SAO~109003 & $-$0.40  &    1.0   & 1.5 &
0.5   &  CH giant &  (18) \\
BD~+75$^\circ$348 & SAO~6630 & $-$0.80  &    1.4   & 1.7    & 0.3
& Extr. C giant   &   (19) \\
\hline
\end{tabular}
\end{center}
\end{table}

\newpage

\begin{table}
\begin{center}
{{\bf {\large T}ABLE 2d.} {\large E}XTRINSIC {\large
G}ALACTIC-HALO {\large AGB} {\large S}TARS}
\\[0.5cm]
\begin{tabular}{|l|l|r|r|r|r|c|l|}
\hline
STAR & ALIAS & [Fe/H] &[ls/Fe] & [hs/Fe]& [hs/ls]& Type & Reference \\
\hline HD~187861    & CGCS~4524 &  $-$1.65  &    1.0   &     1.9 & 0.9 & CH giant
&  (18, 20) \\
HD~198269    & SAO~106516 &  $-$1.40  &    0.8   &     1.2    &
0.4   & idem &  (18) \\
HD~201626    & SAO~89499 &  $-$1.30  &    1.1   &     1.6    & 0.5
& idem &  (18) \\
HD~209621    & HP~Peg &  $-$0.90  &    1.1   & 2.4    &     1.3
& idem &  (18) \\
HD~224959 & BD~-03$^\circ$5751 &  $-$1.60  &    0.8   &     1.9
&     1.1 & idem &  (18) \\
CD~$-$38$^\circ$2151 & SAO~196068 & $-$1.40  &    0.6   &     1.1
&     0.5 & idem &  (18) \\
\hline BD~+67$^\circ$922 & AG~Dra       &       $-$1.30  &    0.5
&
1.0    & 0.5 & Yellow Symb. & (21) \\
BD~$-$21$^\circ$3873 & IV~Vir  & $-$1.32  &    0.4   & 0.5    &
0.1   & idem & (22) \\
He~2-467     &  --- &    $-$1.11  &    1.0   & 1.1    &     0.1
&
idem & (23) \\
\hline HD~196944  & SAO~144688 &   $-$2.45  &    0.6   & 1.3 &
0.7   & C-rich giant  &   (24) \\
LP~625$-$44    & --- &  $-$2.71  &    1.1   & 2.4    &     1.3 &
C-rich subgiant&   (25) \\
CS~22898$-$027 & ---  &   $-$2.35  &    1.0   & 2.4    &     1.4 &
idem & (26, 27) \\
LP~706$-$7     & --- & $-$2.74  &    0.3     &   1.9      &   1.6
&  idem &  (28) \\
\hline HD~74000     & SAO~154538  &    $-$2.08  &    0.3   & 0.3
&     0.0   & N-rich dwarf & (28) \\
HD~25329     & SAO~56928    &  $-$1.84  &    0.4   & 0.5    & 0.1
& idem & (29) \\
\hline
\end{tabular}
\end{center}
(1) Smith \&  Lambert (1985); (2) Smith \& Lambert (1990); (3)
Smith \& Lambert (1986); (4) Abia \& Wallerstein (1998); (5) Van
Winckel   \&  Reyniers (2000); (6) Za\v{c}s et al. (1995); (7)
Reddy et al. (1999); (8) Kipper et al. (1996); (9) Kipper \&
J$\o$rgensen (1994); (10) Tomkin \& Lambert (1983); (11) Smith
(1984); (12) Kov\`{a}cs (1985); (13) Tomkin \& Lambert (1986);
(14) Smith \& Lambert (1984); (15) Tech (1971); (16) Smith \&
Suntzeff (1987); (17) Smith et al. (1993). (18) Vanture (1992);
(19) Za\v{c}s et al. (2000); (20) Vanture (2000); (21) Smith et al
(1996); (22) Smith et al. (1997); (23) Pereira et al. (1998);
(24) Za\v{c}s et al. (1998); (25) Aoki et al. (2000); (26)
McWilliam et al. (1995); (27) McWilliam (1998); (28) Norris et
al. (1997); (29) Beveridge \& Sneden (1994)

\end{table}

\end{document}